\documentclass[twocolumn]{aastex701} %NWM
\usepackage{booktabs}
\usepackage{amsmath}
\usepackage{subcaption} 
\usepackage[percent]{overpic} % for placing text over figures
\usepackage{cleveref}
\usepackage{graphicx}
\usepackage{comment}
\usepackage{multirow}
\usepackage{hyperref}
\usepackage[overload]{textcase}
\hypersetup{
    colorlinks=true,       % True: colored text; False: boxed links
    linkcolor=green,        % Color for internal links (like figures)
    citecolor=blue,       % Color for bibliographic citations
    urlcolor=cyan,          % Color for external URLs
    pagebackref=true     % clickable page number from citations back to the text
}

\newcommand{\obsunits}{\,{\rm K\,km\,s^{-1}\,pc^2}}
\newcommand{\skirt}{\texttt{SKIRT}}
\newcommand{\cloudy}{\texttt{CLOUDY}}
\newcommand{\fire}{\texttt{FIRE}}
\newcommand{\fireTwo}{\texttt{FIRE-2}}
\newcommand{\firebox}{\texttt{FIREbox}}
\newcommand{\core}{\texttt{CORE}}
\newcommand{\massivefire}{\texttt{MassiveFIRE}}
\newcommand{\tripleLatte}{\texttt{Triple Latte}}

\newcommand{\LcoOneZero}{\ensuremath{L'_{{\rm CO}(1-0)}}}
\newcommand{\LcoTwoOne}{\ensuremath{L'_{{\rm CO}(2-1)}}}

\newcommand{\Mstar}{\ensuremath{\mathrm{M}_{\mathrm{star}}}}

\newcommand{\ZgasHl}{\ensuremath{\mathrm{Z}_{\mathrm{gas(R50)}}}}

\newcommand{\Mhalo}{\ensuremath{\mathrm{M}_{\mathrm{halo}}}}
\newcommand{\sfrHundredMyr}{\ensuremath{\mathrm{SFR}_{100\,\mathrm{Myr}}}}
\newcommand{\Mgas}{\ensuremath{\mathrm{M}_{\mathrm{gas}}}}

\newcommand{\RvisibleHl}{\ensuremath{\mathrm{R50}}}

\newcommand{\s}{\ensuremath{\mathrm{s}}}
\newcommand{\g}{\ensuremath{\mathrm{g}}}
\newcommand{\pc}{\ensuremath{\mathrm{pc}}}
\newcommand{\kpc}{\ensuremath{\mathrm{kpc}}}
\newcommand{\ergs}{\ensuremath{\mathrm{erg\,s^{-1}}}}
\newcommand{\Myr}{\ensuremath{\,\mathrm{Myr}}}

\newcommand{\Rs}{\ensuremath{\mathrm{R}_{\mathrm{stop}}}}

\newcommand{\lambdaCDM}{\ensuremath{\Lambda{\rm CDM}}}

\newcommand{\RBFInterpolator}{\texttt{RBFInterpolator}}

\newcommand{\twelvePlusOH}{\ensuremath{12+\log({\rm O/H})}}
\newcommand{\be}{\begin{equation}}
\newcommand{\ee}{\end{equation}}

\defcitealias{Paper1}{Paper~I}
\defcitealias{Paper3}{Paper~III}

\begin{document}
%\doublespace %NWM

\title{Modeling CO Line Luminosity with FIRE simulations: II. Comparison With Observations, CO Luminosity Functions and Redshift Evolution}

\author[0000-0002-3155-946X]{Doğa Tolgay}
\affiliation{Canadian Institute for Theoretical Astrophysics, University of Toronto, 60 St. George Street, Toronto, ON M5S 3H8, Canada}
\affiliation{University of Toronto Physics Department \\
McLennan Physical Laboratories, 60 St George St \\
     Toronto, ON M5S 1A7, CANADA}
\email{doga.tolgay@mail.utoronto.ca}

\author[0000-0003-2358-9949]{J. Richard Bond}
\affiliation{Canadian Institute for Theoretical Astrophysics, University of Toronto, 60 St. George Street, Toronto, ON M5S 3H8, Canada}
\email{bond@cita.utoronto.ca}

% \author[0000-0002-2731-7708]{Nathan Carlson}
% \affiliation{Canadian Institute for Theoretical Astrophysics, University of Toronto, 60 St. George Street, Toronto, ON M5S 3H8, Canada}
% \affiliation{University of Toronto Physics Department \\
% McLennan Physical Laboratories, 60 St George St \\
%      Toronto, ON M5S 1A7, CANADA}
% \email{njcarlson@physics.utoronto.ca}
\author[0000-0002-8659-3729]{Norman W. Murray} 
\affiliation{Canadian Institute for Theoretical Astrophysics, University of Toronto, 60 St. George Street, Toronto, ON M5S 3H8, Canada}
\email{murray@cita.utoronto.ca}

\author[0000-0003-2618-6504]{Dongwoo Chung}
\affiliation{Department of Astronomy, Cornell University, Ithaca, NY 14853, USA}
\email{dongwooc@cornell.edu}

\author[0000-0002-1109-1919]{Robert Feldmann} 
\affiliation{Department of Astrophysics, University of Zurich, Zurich CH-8057, Switzerland}
\email{robert.feldmann@uzh.ch}

\author[0000-0003-0603-8942]{Andrew Wetzel}
\affiliation{Department of Physics \& Astronomy, University of California, Davis, CA 95616, USA}
\email{awetzel@ucdavis.edu}

\begin{abstract}
We present predictions for the CO(1-0) and CO(2-1) luminosities of 765, 850, 887 and 915 galaxies at z=0, 1, 2 and 3, respectively, drawn from the Feedback in Realistic Environments (\fireTwo) simulations and post-processed with \skirt\ radiative transfer and \cloudy\ line emission calculations. We compare our CO line luminosity predictions with existing observations. At z=0, our predictions reproduce the observed scatter and mean relation between star formation rate (SFR) and CO(1-0) luminosity for both massive star-forming galaxies and low-metallicity dwarf galaxies. We also reproduce the observed superlinear power-law slope between CO(1-0) luminosity and metallicity. At z=1, our CO(2-1) predictions scatter to much lower luminosities than observed, which likely reflects the limited sensitivity of current surveys. At fixed SFR, our CO(1-0) luminosities decrease by roughly an order of magnitude per unit increase in redshift. This decline is likely driven by the decrease in metallicity with increasing redshift, since FIRE-2 galaxies with comparable SFR and metallicity produce similar CO(1-0) luminosities regardless of redshift. At fixed stellar mass, the predicted luminosities show no redshift dependence. These results show that our FIRE-2 predictions reproduce the observed scaling relations between CO(1-0) luminosity and galaxy properties wherever the observed galaxies are robustly detected. This gives us confidence in our predictions for the regime where no observational data exist, and suggests that our luminosity predictions can be used to build predictive models for CO line intensity mapping experiments.
\end{abstract}


% \tableofcontents
% \newpage

\section{Introduction} \label{sec:Introduction}

\subsection{CO Line Intensity Mapping \label{sec:intro_lim_LSS}}

Line intensity mapping (LIM) has emerged as a promising alternative to galaxy surveys, probing the large scale structure (LSS). LIM trades angular resolution for mapping speed. Rather than cataloging galaxies, it measures spatial fluctuations in the integrated emission of a spectral line, or lines, from many unresolved sources, including faint objects undetectable in standard galaxy surveys. Since they measure the redshift of emission along the line of sight, LIM experiments yield a three-dimensional tomographic map of the LSS over a broad range of redshifts, at a fraction of total observing time required for galaxy surveys \citep{Kovetz_2017, Kovetz_2019, Bernal_2019}. Many of these lines trace star formation and/or conditions in the interstellar medium (ISM) of galaxies. Thus, LIM surveys provide information about astrophysics as well as cosmology. For example, they could provide a more complete census of star formation across cosmic time than galaxy surveys, e.g., \citet{Visbal2010_LIM, Kovetz_2017, Kovetz_2019, Bernal_2019, Chang_Lidz_2026_LIM}. 

In this series of three papers, we focus on the rotational transitions of carbon monoxide, with an eye toward the CO Mapping Array Project (COMAP, \citealt{Cleary_2022_comap_earlyScience1}), which is currently underway. It targets CO(1-0) at $z\sim3$ and has progressively tightening the upper limits of the CO power spectrum through its first two observing seasons \citep{dongwoo_2022_comap_earlyscience5, dongwoo_2024_COMAP_season2}, but a confirmed detection has yet to made. Interpreting these measurements, first as upper limits and eventually as a detection, requires a physically grounded model of how CO emission from the ISM of galaxies traces the underlying dark matter distribution. This is the second in a series of three papers describing such a model, using state of the art MHD cosmological galaxy simulations from the FIRE project, post-processed with radiative transfer calculations using \skirt\ and emission line calculations using \cloudy, all described briefly in \cite{Paper1} (hereafter \citetalias{Paper1}). In this paper we test the predictions of our model using a number of different observations. \citetalias{Paper3} (in prep.) presents simple analytic models, or response functions, that can be used in combination with large scale N-body or halo models to produce mock maps of CO emission from large volumes of the universe. These mock maps can then be compared quantitatively to maps made by LIM surveys.

\subsection{CO Response Functions \label{sec:intro_empirical_models}}

The critical element for producing mock maps for a blind CO LIM survey is the functional relationship between CO emission and the dark matter field in N-body simulations. We describe this relationship with response functions, $g$, characterized by a mean CO luminosity $\mu_{\log L'_{\mathrm{CO}}}(\Mhalo; z)$ and a scatter $\sigma_{\log L'_{\mathrm{CO}}}(\Mhalo; z)$, where $L'_{CO}$ is the CO luminosity in the units of $\obsunits$ often used by observers \citep{Solomon_1997_ULIRGs_Xco}. Given $g$, line emission can be painted onto the halos of a large-volume dark-matter-only simulations as if the baryonic content of the galaxies had been simulated directly, and the resulting maps used both to forecast cosmic signal strength. Such mock maps are also useful for instrument design, and for testing the analysis pipelines that will be, and are being, applied to real data.

In the current literature these response functions are constructed either from empirical calibrations \citep{Lidz_2011_coModelling, pullen_2013_coModelling, Li_2016_coModelling, Keating_2020_coModelling, dongwoo_2022_comap_earlyscience5} or semi-analytical models of galaxy formation \citep{Gong_2011_coModelling, Lagos2012_Lco_simulatedGalaxies, Sun_2019_coModelling, Yang_2022_coModelling, masRibas2023_LIMfast1, jasonSun2023_LIMfast2_coModelling} which are tuned to match observed scaling laws. The empirical route typically follows a chain of calibrated relationships between halo mass and CO(1-0) luminosity (\LcoOneZero) using a correlation measured in observed galaxies \citep{Tacconi_1987_COvsLir, 1991ARA&A..29..581Y, Carilli_2013, Kamenetzky_2016_herschellSpireArchive_coSLED}. Each link is informed by real data, which is the principal strength of this approach. However, this is also its principal limitation, as the currently available data are dominated by the most luminous galaxies, with CO detections preferentially are from massive, actively star-forming, and metal-rich systems as we discuss below. Luminous galaxies at $z\sim1$ to 3 do appear to follow a relation similar to local ones \citep{Carilli_2013}, but for the faint, low mass galaxies that dominate by number the assumption is untested, because they lie below current sensitivity limits. Partly as a result, published CO models differ by one to two orders of magnitude at the redshifts targeted by COMAP \citep{Breysse_2016_highredshift_COmocks, Breysse_2022_comap_EarlyScience7, dongwoo_2024_COMAP_season2}.

\subsection{Observational Constraints on CO Emission \label{sec:intro_obs_constraints}}

In the local Universe, dwarf galaxies with low metallicity are notoriously faint in CO, and even deep observations with sensitive telescopes detect only a minority of them \citep{Schruba_2012_coLuminositiesDwarfs}. Also see \citet{Krumholz_2011_lowZ_z0_observations} for a compilation of low-metallicity dwarf galaxies. CO has nonetheless been detected in dwarfs with gas phase oxygen abundances of $\twelvePlusOH \approx 7.8$ \citep{Elmegreen_2013_WLM, Rubio_2015_WLM} and even $\approx 7.5$ \citep{Shi_2016_DDO70}, far below the solar value of $8.69 \pm 0.04$ \citep{Asplund2021_MW_metallicity_869}. Across larger samples, the CO detection rate increases with stellar mass, star formation rate, visual extinction ($A_V$), and most strongly with metallicity \citep{Cicone_2017_allsmog}, so observations of metal poor systems frequently yield only upper limits. For a review for CO observations at local Universe see \cite{Saintonge_Catinella_2022_anrev}.

At high redshift, the first CO detections were of ultraluminous infrared galaxies and quasar hosts \citep{Brown1991_IRASF10214_CO, Solomon_1992_IRASF10214_CO, Omont1996_quasar_COdetection}, followed by submillimetre galaxies \citep{Frayer1998_submm_coDetection, Greve_2005_SMG_CO, Tacconi_2008_SMG_CO} and later by massive galaxies on the star forming main sequence \citep{Daddi_2010_BzK_CO, Tacconi_2010_PHIBSS, Tacconi_2013_PHIBSS1}. With ALMA, the molecular gas in the most extreme of these systems can now be resolved on kiloparsec scales, as in hyperluminous infrared galaxies forming stars at up to $\rm \sim1000\,M_\odot\,\mathrm{yr}^{-1}$ \citep{gomez2025_resolvedCO_highZ}. Less massive galaxies have been detected individually mainly when strongly lensed \citep{Baker2004_lensed_starburst, Coppin2007_lensed_lymanBreak_COEmitter, Saintonge_2013_lensed_CO_highZ, DessaugesZavadsky2015_highz_lensedCO}. CO samples remain weighted toward massive, metal rich systems even today. Targeted surveys such as PHIBSS2 sample galaxies on and around the main sequence only above $\log(\Mstar/M_\odot) = 10.1$, where metallicities are close to solar \citep{Freundlich_2019}, and even the flux limited ASPECS survey, which has no preselection, detects mostly galaxies with $\Mstar \gtrsim 10^{10}\,M_\odot$ and solar or super solar metallicities \citep{boogaard_2019_alma_aspecs, Inami_2020_ASPECS_stacking}. For a review of CO observations at high redshift, see \citet{Carilli_2013}.

This selection against lower mass galaxies is important at high redshift, because such galaxies remain metal poor across cosmic time. As we will show, metal poor systems produce substantially less CO emission per unit SFR than their more massive counterparts, so the aggregate contribution of these numerous but CO faint galaxies to the cosmic CO signal remains uncertain, with estimates ranging from tens of percent \citep{Uzgil2019_ALMA_HUDF_CO_shotNoise} to relatively little \citep{Inoue2020_COUniverse}. Current LIM models nonetheless predict this signal by extrapolating from observations of massive galaxies, which is risky for the faint population. There is good evidence that SFR is related to the amount of gas \citep{Kennicutt_1989}, and to molecular gas in particular, \citep{Bigiel2008_KSlaw_molecularGas} but the relation between CO luminosity and gas mass depends strongly on metallicity \citep{Wolfire_2010_coDarkGas, Feldmann2012_Xco_paperI, Madden_2020} and on other local conditions such as the cosmic ray ionization rate \citep{bisbas_2015_co_dark_clouds}.

\subsubsection{Why use hydrodynamic simulations? }

Cosmological (magneto-)hydrodynamic simulation-based calculations offer an alternative to empirical models, and offer a way out of the extrapolation problem. Several studies have already paired cosmological galaxy formation models with a treatment of CO line emission. Because large-volume simulations do not resolve the internal structure of molecular clouds, most of these rely on a subgrid cloud model. \citet{Inoue2020_COUniverse} post-processed \texttt{IllustrisTNG} with a subgrid molecular cloud model and \texttt{DESPOTIC}, reproducing the local CO(1-0) luminosity function, and found that roughly ten per cent of the local molecular gas resides in largely CO-dark dwarfs with $\Mstar<10^9\,M_\odot$. \citet{Garcia2024_slick} similarly treated each gas particle of the $25\,h^{-1}\,{\rm Mpc}$ \texttt{SIMBA} simulation as a subresolution cloud with an assumed power-law density profile, computed its CO emission with \texttt{DESPOTIC}, and used a random forest regression trained on a subset of clouds and galaxies to extend the predictions to the full volume. Their CO(1-0) luminosity functions at $z=0$, 1 and 2 are consistent with xCOLD GASS, ASPECS and COLDz. In both cases, however, the CO emission depends on the assumed structure of clouds that the simulation itself does not resolve.

Higher-resolution simulations such as zoom-in runs with \fireTwo\ physics \citep{Hopkins_2018_fire2} resolve the cold, dense phase of the ISM directly, so the CO emission can in principle be computed from the simulated gas without an assumed cloud structure. \citet{Keating_2020} and \citet{Thompson2024_FIRE_CO} post-processed \fireTwo\ galaxies in this way, using the \texttt{LEBRON} radiation field, \texttt{CHIMES} chemistry and \texttt{RADMC-3D} line radiative transfer. Even so, the CO emission was not straightforward to recover. \citet{Keating_2020} found that the predicted CO(1-0) luminosity depends strongly on the adopted shielding length; the common choices, such as the Jeans length or the smoothing length, overproduce CO, and only a shielding length about an order of magnitude shorter reproduces the Milky Way CO luminosity and $X_{\rm CO}$. This length effectively acts as a tuned subgrid parameter. \citet{Thompson2024_FIRE_CO} showed that relaxing the assumption of chemical equilibrium changes the $\rm H_2$ masses and CO luminosities by $\approx25\%$. Both studies predict excess CO luminosity at a given SFR for $\rm SFR>1\,M_\odot\,{\rm yr}^{-1}$.

We follow the same resolved ISM route, but apply a different post-processing pipeline (\citetalias{Paper1}) to a much larger galaxy sample drawn from several FIRE simulation suites with FIRE-2 physics \citep{Hopkins_2018_fire2}. The cosmological volume \firebox\ \citep{Feldmann_2023} provides a representative, volume limited galaxy population at high resolution, reaching to dwarf galaxies and SFRs below current detection limits. To increase the number of massive halos that hosts the luminous galaxies dominating current CO detections, we further used the \massivefire\ zoom-in simulations \citep{Feldmann2016_massiveFIRE_FIRE1, Feldmann_2017_massiveFIRE, AnglesAlcazar2017_massiveFIRE_FIRE2} at $\rm z=2$ and 3. We also include the \core\ \fireTwo\ simulations with MW- and LMC-mass galaxies \cite{Wetzel_2023, Wetzel2025_secondDataRelease_FIRE2} and a higher resolution re-run of m12i from the \tripleLatte\ suite \citep{Wetzel_tripleLatte}, which let us test whether our \LcoOneZero\ predictions depend on resolution. For each galaxy we compute the radiation field with three dimensional dust radiative transfer in \skirt\ \citep{Camps_and_baes_2020_skirt9} and the chemistry and CO line emission with \cloudy\ \citep{cloudy_version_c25_release}, using the density, metallicity and turbulence of the simulated gas.

An obvious objection to this approach of making prediction is that such predictions may simply be wrong. A simulation is only as good as its treatment of star formation and feedback, and a post-processing pipeline introduces further assumptions about the simulated clouds structure, shielding, and the interpolation of a finite grid of models onto gas properties and the consequent line emission.

Unlike \citet{Keating_2020}, we calibrated neither the simulation nor the pipeline in this study. Because nothing in the calculation has been tuned to reproduce CO measurements, the comparison of our predictions against existing observations constitutes a genuine test rather than a consistency check. The agreement, if and where it occurs, can be taken as evidence that the underlying post-processing is sound, and that the predictions outside the range of current observations can be trusted. Accordingly, in this paper we compare our CO(1-0) and CO(2-1) luminosity predictions across a broad range of redshifts and galaxy properties using current data.

This paper is organized as follows. In Sections~\ref{sec:Simulation_Methodology} and \ref{sec:post_processing_fire} we summarize the simulations and the post-processing pipeline. In Section~\ref{sec:observational_data} we describe the observational datasets against which we compare. In Section~\ref{sec:results_and_comparison_with_observations} we present the comparison, covering the \LcoOneZero-SFR relation and its metallicity dependence at $z=0$, the local CO(1-0) luminosity function, the \LcoTwoOne-SFR relation at $z=0$ and $z\sim1$, the evolution of \LcoOneZero\ with redshift, and the CO(1-0) luminosity function at $z\geq2$. In Section~\ref{sec:discussion} we assess what these comparisons do and do not establish, and set out the systematic uncertainties affecting our predictions. We conclude in Section~\ref{sec:conclusion}.

\section{Simulation Methodology \label{sec:Simulation_Methodology}}
 
Our galaxy sample and the pipeline used to compute its line emission are described in full in \citetalias{Paper1}. Here we summarize only the elements need to interpret the results presented below.
 
We use galaxies drawn from four suites of simulations run with the \texttt{GIZMO} code \citep{Hopkins_2015} as part of the Feedback in Realistic Environments (\fireTwo) project. These are the \core\ \fireTwo\ suite \citep{Hopkins_2018_fire2}, a new higher resolution version of m12i as a part of the new \tripleLatte\ suite \citep{Wetzel_tripleLatte}, the \firebox\ cosmological volume \citep{Feldmann_2023}, and \massivefire\ \citep{Feldmann2016_massiveFIRE_FIRE1, Feldmann_2017_massiveFIRE, AnglesAlcazar2017_massiveFIRE_FIRE2}. All of the simulation suites used in this work adopt the \fireTwo\ physics model with the meshless finite-mass (MFM) hydrodynamic solver \citep{Hopkins_2018_fire2}. All runs follow radiative heating and cooling self-consistently, explicitly track 11 elements, and form star particles only from gas that is dense, predominantly molecular, and locally self-gravitating, and Jeans unstable \citep{Hopkins_2018_fire2}. Stellar feedback includes photoionization, photoelectric heating, radiation pressure, winds, and supernovae, with energy and momentum budgets set by a \cite{Kroupa_2001} IMF coupled to the STARBURST99 population synthesis model \citep{Leitherer_1999}. None of the runs include feedback from active galactic nuclei. 

The cosmological parameters vary slightly across the simulation suites. The \core\ \fireTwo\ halos were run using either the AGORA cosmology \citep{Kim2014_agoraCosmology_FIRE} ($\Omega_{m} = 0.272$, $\Omega_{\Lambda} = 0.728$, $\Omega_{b} = 0.0455$, $h = 0.702$, $\sigma_{8} = 0.807$, $n_{s} = 0.961$; m11q, m12i, m12f, m12m, m12b, m12c) or the Planck 2013 cosmology \citep{Planck_2014} ($\Omega_{m} = 0.31$, $\Omega_{\Lambda} = 0.69$, $\Omega_{b} = 0.048$, $h = 0.68$, $\sigma_{8} = 0.82$, $n_{s} = 0.97$; m11d, m11e, m11h, m11i, m12r, m12w) \citep{Hopkins_2018_fire2, ElBadry2018_m11halosPlanck}. FIREbox uses the cosmological parameters from \cite{Planck_2016} ($\Omega_{m} = 0.3089$, $\Omega_{\Lambda} = 0.6911$, $\Omega_{b} = 0.0486$, $h = 0.6774$, $\sigma_{8} = 0.8159$, $n_{s} = 0.9667$), whereas the MassiveFIRE halos \citep{Feldmann2016_massiveFIRE_FIRE1, Feldmann_2017_massiveFIRE, AnglesAlcazar2017_massiveFIRE_FIRE2} use $\Omega_{m} = 0.2821$, $\Omega_{\Lambda} = 0.7179$, $\Omega_{b} = 0.0461$, $h = 0.697$, $\sigma_{8} = 0.817$, and $n_{s} = 0.9646$. All simulations assume a flat $\Lambda$CDM cosmology with $w = -1$.

The four suites are complementary in resolution and dynamic range. The \core\ \fireTwo\ runs consists of seven Milky Way mass m12 halos with halo masses approximately around $\rm10^{12}\,M_\odot$ and five lower mass m11 halos with halo masses approximately several $\rm 10^{11}\, M_\odot$. All of our \core\ \fireTwo\ runs have a baryonic particle mass resolution of $7100\,M_\odot$, with fully adaptive gas softening that reaches $1\,\pc$ in the densest regions. We further included one \tripleLatte\ version of m12i galaxy with improved baryonic mass resolution by a factor of eight ($875\,M_\odot$) and allows us to test the resolution dependence of our predictions.

\firebox\ simulation is a high resolution hydro cosmological box simulation while resolving the ISM. Its volume provides a statistically representative population of galaxies over a wide range of masses. The initial condition selection for the \firebox\ is weighted towards box without an exceptionally massive halo at $z=0$, thereby maintaining a computational efficiency.

By number, the \firebox\ sample is dominated by dwarf galaxies. The same is also true for our Universe, however, these galaxies are often too faint to be detected by current CO surveys. Albeit, they are naturally included in the simulation and establish the faint end of the luminosity functions we present here, while otherwise it is difficult to probe with observational surveys. The \firebox\ volume is particularly well suited to studying typical galaxy populations and their ISM, whereas the rarest, most massive cluster-scale halos ($\rm M_{\rm halo}\approx10^{14}$--$10^{15}\,M_\odot$) are not represented. At $z=3$, the most massive halo in \firebox\ has $\rm M_{\rm halo}\approx1.1\times10^{12}\,M_\odot$. To extend the sampling toward the high-mass end of the halo mass function, we additionally use snapshots from the \massivefire\ suite. With \massivefire, the maximum halo mass at $z=3$ reaches $\rm M_{\rm halo}\approx1.6\times10^{12}\,M_\odot$. The \massivefire\ simulations ran down to z=1, however, we used the four snapshots at $z=2$ and 3 each. We note that at $z=3$ the virial mass of the \massivefire\ halos are still less massive than the extreme high mass tail halos at flat \lambdaCDM\ cosmology. 

To determine the halo masses of our \fire\ suites, we used the \texttt{ROCKSTAR} halo finder \citep{Behroozi_2013} for the \core, \tripleLatte\ and \massivefire\ simulations, and the \texttt{AMIGA} Halo Finder \citep{Gill_2004, Knollmann_2009} for the \firebox\ cosmological volume. These halo finders had already been run by the developers of the \fire\ suites, so we used their output directly. For both halo finders we adopt the virial mass definition. The resulting halo masses are therefore lower than the $\rm M_{200m}$ values reported for the \core\ \fireTwo\ galaxies \citep{Wetzel_2023, Wetzel2025_secondDataRelease_FIRE2}.

As described in \citetalias{Paper1}, we use a catalog containing 765, 850, 887, and 915 galaxies at $z=0$, 1, 2, and 3, covering a wide  range in galactic properties: $6 < \log(\Mstar/M_\odot) < 11.6$, $10.1 < \log(\Mhalo/M_\odot) < 13.2$, $7.9 < \log(\Mgas/M_\odot) < 10.8$, $-3.3 < \log(\sfrHundredMyr/(M_\odot,\mathrm{yr}^{-1})) < 2.2$, and $-2.2 < \log(\ZgasHl/Z_\odot) < 0.9$. Gas particle metallicities in \fireTwo\ are stored as mass fractions, which we normalise to solar using a fiducial value of 0.0134 \citep{Grevesse2010_solarMetallicity}.

\section{Calculating Line Emission from FIRE Snapshots \label{sec:post_processing_fire}}
 
Line luminosities are computed in post-processing by combining \skirt\ (to calculate the local radiation field) with \cloudy\ (to calculate the line emission and abundances given the local radiation and density fields). The pipeline returns $\mathrm{CO(1-0)}$ and $\mathrm{CO(2-1)}$ luminosities for every gas particle. We refer the reader to \citetalias{Paper1} for the full description of the pipeline and summarize here only the elements needed to interpret the results below.

We first use the open-source 3D Monte Carlo dust radiative transfer code \skirt\ (version 9, \citealt{Baes_2011, Baes_2015, Camps_2015, Camps_and_baes_2020_skirt9}) to compute the interstellar radiation field (ISRF) that each gas particle is exposed to. Star particles are treated as single stellar populations from STARBURST99 \citep{Leitherer_1999} as a function of age, metallicity, and initial mass. We assume a constant dust to metal ratio $\delta_{dzr}=0.4$, consistent with the \fireTwo\ simulations \citep{Hopkins_2018_fire2}, except for gas hotter than $10^{6}\,{\rm K}$, which we assume to be dust free as a result of thermal sputtering \citep{Tielens_1994}, and adopt the \cite{Weingartner_2001} dust mix with $R_v = 3.1$. Galaxies are gridded onto $10^5$ Voronoi tessellation cells and illuminated with $10^7$ photon packages for each of 350 logarithmically spaced wavelength bins spanning $\lambda = 0.03-2000\,\mu m$. The far-ultraviolet (6-13.6 eV) energy density in each cell, which controls both photoelectric heating and CO photodissociation \citep{Draine_2011}, is obtained by integrating the \skirt\ mean intensity $J_\lambda$ and normalizing to the Habing field \citep{Habing1968_radiationFieldMeasurement} and the MW value of \cite{Draine_1978_ISRF_MW_value_habingUnits} to give $G_{\rm FUV}$.

Line emission is then computed with version 25.00 of \cloudy, last described by \cite{cloudy_version_c25_release}, which predicts level populations and line emission for atoms and molecules using the Leiden Atomic and Molecular Database (LAMDA, \citealt{Schoier_2005}). Running \cloudy\ for every gas particle is prohibitively expensive, so we instead construct a lookup table of luminosity per unit gas mass spanning the five physical quantities that control the emission: metallicity ($-3 \leq \log(Z_{gas}/Z_\odot) \leq 1$), hydrogen number density ($-4 \leq \log(n_H/cm^{-3}) \leq 3.5$), turbulent velocity ($0 \leq \log(v_{gas}/kms^{-1}) \leq 2$), ISRF ($-5 \leq \log(G_{\rm FUV}/G_0) \leq 5$), and stopping length ($1 \leq \log(h/pc) \leq 3.5$) at a single redshift. Because our gas cells never exceed  $n_{\rm H}R_{\rm stop} < 10^{5.6}\,{\rm cm^{-3}\,pc}$ we only use the \cloudy\ models below this column density. At a single redshift we ran $27{,}324$ \cloudy\ models, however, we repeat the same models at different CMB temperatures appropriate for z=0, 1, 2 and 3. Thus, in total we ran $4\times27{,}324=109{,}296$ \cloudy\ models of which approximately 0.6\% fail to complete.

Turbulent velocities are measured directly from the \fireTwo\ simulations as the magnitude of each particle's velocity relative to the kernel-weighted center of mass velocity of its neighbors within two smoothing lengths, computed using \texttt{Meshoid}\footnote{\url{https://github.com/mikegrudic/meshoid}}.Our definition of the turbulent velocities include rotation, shear, shocks and bulk motions rather than unresolved microturbulence. The simulation is allowed to continue until each \cloudy\ model reaches to a certain stopping length $\Rs$. This value is selected to cover the smoothing length of gas cells in our \fireTwo\ runs.

Each model adopts an plane-parallel (an open geometry in \cloudy\ terminology), constant-density geometry illuminated externally by a scaled MW radiation field measured by \citet{Black_1987_isrf}. ISM abundances are metal and grain abundances are first set to their typical diffuse ISM value and then scaled with metallicity that covers the metallicities of gas cells at fixed grain to metal ratio. We normalized the mass ratios given in FIRE snapshots by $\rm Z_\odot=0.0134$ \citep{Grevesse2010_solarMetallicity} and we use this value for scaling. That is, we did not use the carbon and oxygen abundances of individual gas cells in our \cloudy\ models. Including them would require an extra parameter in our \cloudy\ tables, which would make the computation much more expensive. Our predicted CO luminosities are sensitive to assumed normalization factor of $\rm Z_\odot=0.0134$.

Cosmic rays at the average MW background of $2\times10^{-16}\,s^{-1}$ \citep{Spitzer_1968, Ferland_1984, Ferland_2009}. The runs are iterated to convergence to capture the high optical depths reached by the CO lines deep inside molecular clouds.

For each zone of the plane-parallel slab, \cloudy\ prints the angle-averaged, frequency-integrated emissivity $\epsilon_{line} = 4\pi j_{line}$, from which we obtain the luminosity per unit gas mass of every line we track:
\begin{equation}
    \frac{L_{\mathrm{line}}}{M}
    =
    \frac{4\pi}{\rho \Rs} \int_{0}^{\Rs} j_{\mathrm{line}}(s)\, ds \, (\ergs M_\odot^{-1}),
    \label{eq:luminosity_per_mass_calculation}
\end{equation}
where $\rho$ is the gas density, assumed independent of depth $s$ within each \cloudy\ run and $\Rs$ is the stopping length of our \cloudy\ models. The motivation behind this calculation is explained in detail in \citetalias{Paper1}.

Finally, we interpolate the tabulated luminosity per unit mass onto every gas particle in the host galaxy using the \RBFInterpolator\ implemented in the \texttt{SCIPY} \texttt{Python} library \citep{scipy_2020}, trained on the 200 nearest neighbours in the parameter space of the \cloudy\ table, and fall back to the \texttt{NearestNDInterpolator} in the rare cases requiring extrapolation. Multiplying by the particle gas mass and summing over the galaxy gives the CO(1-0) and CO(2-1) luminosities in units of $\ergs$. 

Then we converted the units from $\ergs$ to commonly used by observers, $\obsunits$, following \citet{Carilli_2013}:
\begin{equation}
    \label{eq:co_emission_unitConversion}
    L'_{CO} = \frac{L_{CO}(\ergs)}{L_\odot} \times \frac{1}{3 \times 10^{-11} \times (\nu_{J}/{\rm GHz})^3},
\end{equation}
where $\nu_{J}$ is the rest frequency of the corresponding CO emission. 

\section{Observational data \label{sec:observational_data}}

In Section~\ref{sec:results_and_comparison_with_observations} we compare our simulation results with several observational surveys, relating CO emission to galactic star formation rates (SFR), stellar masses (\Mstar), and metallicities ($\rm Z$ or $\twelvePlusOH$). The surveys are summarized in Table~\ref{table:observational_data}. xCOLD GASS, PHIBSS2, and ASPECS are also used in \citetalias{Paper1} to test our predicted molecular gas masses; here we describe only the properties relevant to the CO luminosity comparisons. These are the observed transition, the sensitivity limit, and how metallicities were inferred.

\begin{table*}
\centering
\small
\setlength{\tabcolsep}{2pt}
\caption{Observational datasets used in this paper. $J_{\rm up}$ lists the upper levels of the observed CO transitions. $\log L'_{\rm lim}$ is the approximate sensitivity limit or the faintest detection, in $\obsunits$, for the CO transition given in parentheses. Metallicities are given as $12+\log({\rm O/H})$.}
\label{table:observational_data}
\begin{tabular}{lccccccccl}
\toprule
Survey & $J_{\rm up}$ & $z$ & $N_{\rm gal}$ & $\log\Mstar$ & SFR & $\log L'_{\rm lim}$ & Metallicity & Figures & Reference \\
 & & & & $[M_\odot]$ & $[M_\odot\,{\rm yr}^{-1}]$ & & & & \\
\midrule
xCOLD GASS & 1, 2 & 0.01--0.05 & 532 & 9--11.3$^a$ & $\approx$0--35$^a$ & $\sim$8.6-7.5 (1-0) & 8.2--8.9$^a$ & 
\ref{fig:Lco10_vs_sfr10Myr_galaxyType_z0_colorBarLogOH},
\ref{fig:co10_lf_z0_observations}, \ref{fig:Lco21_vs_sfr10Myr_galaxyType_z0},
\ref{fig:Lco_10_z0_z1_z2_z3_sfr100Myr},
\ref{fig:Lco_vs_Mstar_redshifts}& \citealt{Saintonge_2017} \\
KLM & 1 & $\approx$0 & 53 & 8.0--10.8$^b$ & 0.01--20 & 5.1 (1-0)$^c$ & 8.0--9.0 & \ref{fig:Lco10_vs_sfr10Myr_galaxyType_z0_colorBarLogOH},
\ref{fig:Lco_10_z0_z1_z2_z3_sfr100Myr},
\ref{fig:Lco_vs_Mstar_redshifts}& \citealt{Krumholz_2011_lowZ_z0_observations} \\
ALLSMOG & 2 & 0.01--0.03 & 88 & 8.5--10 & 0.04--12 & 7.6 (2-1) & 8.2--8.9 & \ref{fig:Lco21_vs_sfr10Myr_galaxyType_z0} & \citealt{Cicone_2017_allsmog} \\
PHIBSS2 & 2 & 0.5--0.8 & 61 & 10.0--11.6 & 3.5--80 & 8.6 (2-1)$^c$ & 8.59--8.74$^d$ & \ref{fig:Lco21_vs_sfr10Myr_galaxyType_z0}, \ref{fig:Lco21_vs_sfr100Myr_galaxyType_z1_matchedMstarSFR}
\ref{fig:matched_fgas_tdep_vs_Mstar_z1_sfr100Myr_Mstar0.2_fgas0.2_tdep0.2},
\ref{fig:matched_Lco21_vs_sfr100Myr_z1_Mstar0.2_fgas0.2_tdep0.2},& \citealt{Freundlich_2019} \\
ASPECS & 2--4 & 1.0--3.6 & 18 & 9.7--11.5 & 8--318 & 9.0 (2-1) & 8.70--9.05 & \ref{fig:Lco21_vs_sfr100Myr_galaxyType_z1_matchedMstarSFR},
\ref{fig:matched_fgas_tdep_vs_Mstar_z1_sfr100Myr_Mstar0.2_fgas0.2_tdep0.2},
\ref{fig:matched_Lco21_vs_sfr100Myr_z1_Mstar0.2_fgas0.2_tdep0.2}& \citealt{boogaard_2019_alma_aspecs} \\
COLDz & 1, 2 & 1.95--2.85 (1-0) & --- & --- & --- & 9.5 (1-0) & --- & \ref{fig:co10_lf_z23_observations} & \citealt{Riechers_2019_coldz} \\
 & & 4.9--6.7 (2-1) & & & & & & & \\
\bottomrule
\end{tabular}
\par\smallskip
\raggedright\footnotesize
$^a$ Extracted from the publicly available xCOLD GASS catalog. \\
$^b$ Compiled in this work from the literature; individual values and references are listed in Table~\ref{table:CO_SFR_metallicity_krumholz2011}. \\
$^c$ Faintest detection. \\
$^d$ Not measured; inferred from the stellar mass--metallicity relation of \cite{Genzel_2015} on the \cite{pettini_pagel_2004_12+logOH_prediction} scale. Range computed here by applying this relation to the stellar masses and redshifts of the 61 galaxies. \\
\end{table*}

\subsection{Low redshift $(0\lesssim z\lesssim 1)$ surveys}

For CO(1-0) emission from individual galaxies at $z\approx0$, we use the xCOLD GASS survey \citep{Saintonge_2017} and the data compilation of \citet{Krumholz_2011_lowZ_z0_observations}. xCOLD GASS is one of the most comprehensive surveys of molecular gas in nearby galaxies, providing IRAM 30m CO(1-0) observations for 532 stellar-mass-selected galaxies with $\Mstar>10^9\,M_\odot$ over the redshift range $0.01<z<0.05$. A subset of the galaxies also has CO(2-1) observations. The sample spans stellar masses of $9<\log(\Mstar/M_\odot)<11.3$, star formation rates up to $\sim35\,M_\odot\,\mathrm{yr}^{-1}$, and gas-phase metallicities of $8.2<12+\log({\rm O/H})<8.9$, derived with the O3N2 calibration of \citet{pettini_pagel_2004_12+logOH_prediction} for star-forming galaxies and from the mass--metallicity relation \citep{kewleyMetallicityCalibrationsMassMetallicity2008} otherwise. The xCOLD GASS strategy was to observe each galaxy until \LcoOneZero\ was detected with $\rm S/N > 5$ or the rms noise was low enough to allow $\rm 5\sigma$ upper limits on the gas fraction of $\rm M_{H_2}/\Mstar = 1.5\%$ for COLD GASS and $2.5\%$ for COLD GASS low. Because there is no luminosity cut, and the galaxies are instead observed until a fixed gas mass fraction is reached, the depth in terms of CO luminosity varies by about an order of magnitude across the stellar mass range of the sample. At the same time, this limit is formally defined in terms of the molecular gas mass fraction, which is inferred from the CO emission by assuming the \cite{Accurso_2017} conversion factor, and it therefore implies a sensitivity limit on the CO luminosity as a function of stellar mass. At the high stellar mass end, this limit corresponds to $\log(\LcoOneZero)\sim8.6$, while at the low stellar mass end the sensitivity limit is $\log(\LcoOneZero)\sim7.5$, which they report as the completeness limit of the survey resulting from the stellar mass cut. We also compare the \firebox\ CO(1-0) luminosity functions at $z=0$ with those from the xCOLD GASS survey.

We also use the galaxy-integrated CO luminosities, SFRs and metallicities of 53 galaxies compiled by \citet{Krumholz_2011_lowZ_z0_observations} (hereafter KLM). The sample includes both dwarf and spiral galaxies and spans $8.0\leq12+\log({\rm O/H})\leq9.0$. The metallicities are drawn from several literature sources with different strong-line calibrations, so their overall normalization is uncertain by $\sim0.1$-$0.2$ dex. We note that solar metallicity corresponds to $12+\log({\rm O/H})\approx8.7$-$8.8$ depending on the calibration (KLM adopt $8.76$). Therefore their compilation includes low metallicity galaxies than xCOLD GASS sample detects. The surveys compiled by KLM reach much lower \LcoOneZero\ than xCOLD GASS (down to $\LcoOneZero\approx10^{5}\,\obsunits$), making this our primary comparison sample for metal-poor galaxies. The \citet{Krumholz_2011} compiled version of the table does not contain stellar masses of these galaxies. We have compiled the stellar masses of these galaxies from the literature and tabulate them in Table~\ref{table:CO_SFR_metallicity_krumholz2011} in Appendix~\ref{sec:krumholzCompiledData}.

For the CO(2-1) transition, we compare with the CO(2-1) observations of xCOLD GASS sources, the APEX low-redshift legacy survey for molecular gas \citep[ALLSMOG;][]{Cicone_2017_allsmog}, and the PHIBSS2 survey \citep{Freundlich_2019}. ALLSMOG reports CO(2-1) observations of 88 nearby ($0.01<z<0.03$) low-mass ($10^{8.5}<\Mstar/M_\odot<10^{10}$) star-forming galaxies obtained with the APEX telescope. The metallicities of the galaxies are inferred using five different calibration methods following \cite{Tremonti_2004_MZR} and \cite{pettini_pagel_2004_12+logOH_prediction}. At their sensitivity limit ($\LcoTwoOne\gtrsim4\times10^{7}\,\obsunits$), the CO detection rate of the full 97-galaxy sample (including nine galaxies observed with the IRAM 30\,m telescope) is 47\%. Non-detections occur predominantly in galaxies with low stellar mass, low SFR, and low gas-phase metallicity.

The PHIBSS2 survey provides CO(2-1) observations of 61 galaxies at $0.5<z<0.8$ (60 detections), obtained with the IRAM Plateau de Bure Interferometer and its NOEMA upgrade, spanning $10.0\lesssim\log(\Mstar/M_\odot)\lesssim11.6$ and star formation rates of approximately $3.5$--$80\,M_\odot\,\mathrm{yr}^{-1}$. The faintest detection has $\LcoTwoOne\approx10^{8.6}\,\obsunits$. The metallicities of the galaxies are inferred from the stellar mass--metallicity relation adopted by \citet{Genzel_2015}, and the inferred metallicities close to solar.

\subsection{High redshift surveys}

At higher redshifts, we use the ASPECS survey \citep{boogaard_2019_alma_aspecs} and the COLDz survey \citep{Riechers_2019_coldz}. ASPECS provides CO detections for galaxies in the Hubble Ultra Deep Field over $1\lesssim z\lesssim3.6$, including CO(2-1) emitters at $1<z<2$, CO(3-2) emitters at $2<z<3$, and a CO(4-3) emitter at $z\approx3.6$. The detected galaxies span SFRs of approximately $8$--$318\,M_\odot\,\mathrm{yr}^{-1}$ and stellar masses of $9.7<\log(\Mstar/M_\odot)<11.5$. There is only four have metallicity measurements available. The inferred metallicities are solar or supersolar. The sensitivity limit of the survey depends on the depth of the survey. For example, $M_{\rm H_2} \geq 10^{9.7}\,M_\odot$ at $1.0<z<1.7$ for a line FWHM of $100\,{\rm km\,s^{-1}}$, and $M_{\rm H_2} \geq 10^{10.3}\,M_\odot$ at $2.0<z<3.1$ for $300\,{\rm km\,s^{-1}}$. We convert these limits by using their assumed  $\alpha_{\rm CO}=3.6\,M_\odot\,(\obsunits)^{-1}$, which yields $\LcoOneZero \approx 10^{9.1}$ and $10^{9.7}\,\obsunits$ at $1.0<z<1.7$ and $2.0<z<3.1$.

COLDz provides blind measurements of the CO(1-0) luminosity function at $2\lesssim z\lesssim3$ with the Karl G. Jansky Very Large Array, together with constraints on the CO(2-1) luminosity function at $5\lesssim z\lesssim7$, covering comoving volumes of $151{,}231\,\mathrm{Mpc}^3$ and $223{,}684\,\mathrm{Mpc}^3$ for CO(1-0) and CO(2-1), respectively. Their measurements reach a luminosity limit of $\log(\LcoOneZero/\obsunits)\approx9.5$. We compare the \firebox\ CO(1-0) luminosity functions at $z=2$ and $z=3$ with those COLDz CO(1-0) measurements.

\section{Results and Comparison with Observations} \label{sec:results_and_comparison_with_observations}

In \citetalias{Paper1} we compared our predictions for CO luminosities of Milky Way molecular clouds with observations. Here we move to the scale of whole galaxies, comparing our predicted CO(1-0) and CO(2-1) luminosities as functions of star formation rate, stellar mass and metallicity to observations of the same quantities at several redshifts. We use the values of SFR, \Mstar, and metallicity ($Z$) directly from the simulation, and \LcoOneZero\ from our \cloudy\ interpolations. Following \citetalias{Paper1}, each galaxy is rotated so that the perpendicular axis is aligned with the net angular momentum of its gas, and we use all particles within $\rm 20\,\kpc$ of the halo center, further restricted the perpendicular axist to $d_{\perp}<5\,\kpc$ of the disc plane for galaxies with $\rm \Mhalo>3\times10^{11}\,M_\odot$ at $\rm z=0$ and all of the galactic properties are measured withing the indicated limits. 

\Mstar\ is the total stellar mass of the galaxy, obtained by summing the masses of all star particles in this region. The \sfrHundredMyr\ is the total mass of the star particles formed within the past $100\Myr$, divided by $100\Myr$. For the \fireTwo\ galaxies, $Z$ always refers to \ZgasHl, the mass-weighted mean metallicity of the gas particles within the half-light radius \RvisibleHl, where the gas particle metallicities are the metal mass fractions stored in the \fireTwo\ snapshots, normalized to $Z_\odot=0.0134$ \citep{Grevesse2010_solarMetallicity}. \RvisibleHl\ is measured from the face-on \skirt\ image of each galaxy, which includes dust attenuation and scattering. We integrate the surface brightness of each pixel over the visible band ($400\leq\lambda\leq700\,{\rm nm}$), sum the resulting luminosities in concentric annuli centered on the galaxy, and take \RvisibleHl\ as the projected radius at which the cumulative luminosity first exceeds half of the total luminosity in the image. We also compute the oxygen gas phase abundances \twelvePlusOH\ from the same gas cells within \RvisibleHl, by using the abundances oxygen abundances carried by each gas cell in the simulation.

We then compare to the observed value of \LcoOneZero\ for the values of SFR, \Mstar, and metallicity inferred from astronomical observations, where the metallicity is reported as \twelvePlusOH\ derived from emission-line ratios. Thus, in this section we are not making an apples-to-apples comparison.

\subsection{ $L'_{CO(1-0)}$  versus SFR at $z=0$ and it's dependence on metallicity \label{sec:Lco10_sfr_metallicity_at_z0}}

\begin{figure*}[ht!]
    \centering
\includegraphics[width=0.99\linewidth]{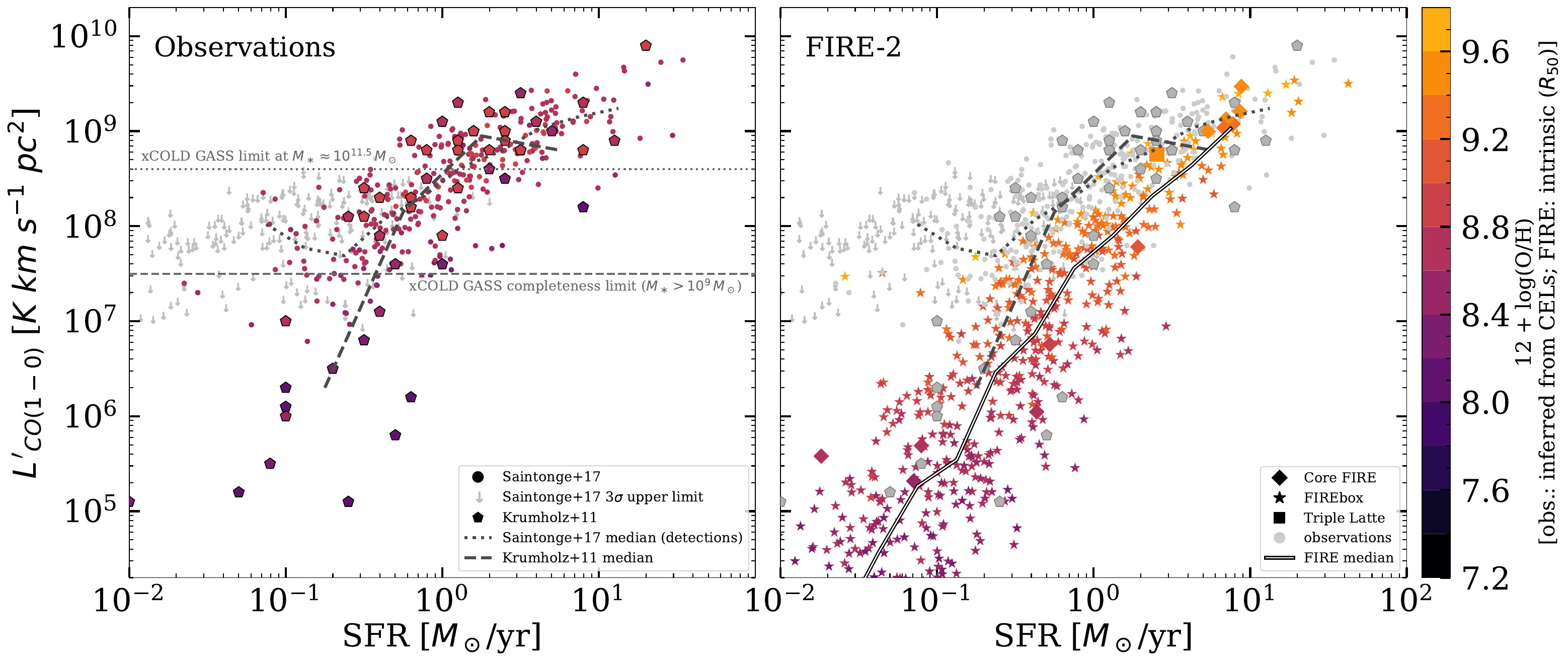}
    \caption{\LcoOneZero\ as a function of SFR, color-coded by gas-phase oxygen abundance, for observed galaxies (left) and post-processed \fireTwo\ galaxies (right). \textit{Left:} xCOLD GASS detections (circles) and the KLM compilation (pentagons), colored by their \twelvePlusOH. xCOLD GASS $3\sigma$ upper limits are shown as grey arrows. The horizontal dashed grey line marks $\log(\LcoOneZero)=7.5$, below which the xCOLD GASS survey \citep{Saintonge_2017} becomes incomplete due to its stellar mass cut ($\Mstar > 10^{9}\,M_\odot$), while the horizontal dotted grey line marks $\log(\LcoOneZero)\approx 8.6$, the gas fraction limited CO sensitivity of xCOLD GASS for its most massive galaxies ($\Mstar \approx 10^{11.5}\,M_\odot$). \textit{Right:} \firebox\ galaxies (stars), \core\ \fireTwo\ galaxies (diamonds) and \tripleLatte\ rerun of the m12i galaxy (square), colored by their \twelvePlusOH. The observations from the left panel are repeated as light grey symbols for reference. For the \fireTwo\ galaxies, the SFR is the total stellar mass formed within the past $100\Myr$, divided by $100\Myr$ (\sfrHundredMyr). Both panels share the color bar, which is divided into steps of 0.2\,dex. For the \fireTwo\ galaxies, \twelvePlusOH\ is calculated from the intrinsic oxygen and hydrogen abundances of the gas cells within the half-light radius, \RvisibleHl. In contrast, the oxygen abundances of the observed galaxies are inferred from collisionally excited lines (CELs). Because different CEL calibrations are known to disagree with one another \citep{Tremonti_2004_MZR, kewleyMetallicityCalibrationsMassMetallicity2008, Moustakas_2010_krumholzCompiledData}, systematic offsets can arise both between different observational samples and between observations and simulations. The lines show the median $\log \LcoOneZero$ in bins of $\log {\rm SFR}$. All \fireTwo\ galaxies combined shown with solid black line on the right panel, xCOLD GASS detections shown with dotted line and KLM with dashed line. The \fireTwo\ and xCOLD GASS medians use 0.25\,dex bins containing at least five galaxies. The KLM median uses 0.5\,dex bins containing at least three galaxies. The xCOLD GASS median excludes upper limits and is therefore biased high at ${\rm SFR} \lesssim 0.3\,M_\odot\,{\rm yr}^{-1}$. \label{fig:Lco10_vs_sfr10Myr_galaxyType_z0_colorBarLogOH}}
\end{figure*}

Figure~\ref{fig:Lco10_vs_sfr10Myr_galaxyType_z0_colorBarLogOH} shows \LcoOneZero\ at $z=0$ as a function of SFR, with the color scale indicating the gas-phase oxygen abundance, \twelvePlusOH. The left panel shows the observations and the right panel the \fireTwo\ galaxies. The xCOLD GASS detections are plotted as filled circles and its $3\sigma$ upper limits as grey downward-pointing arrows. We also show the compilation of nearby galaxies by \citet{Krumholz_2011_lowZ_z0_observations}, which we refer to as KLM, as pentagons. The KLM compilation reaches much lower \LcoOneZero\ than the xCOLD GASS survey and shows a larger scatter. In the right panel, \firebox\ galaxies are shown as stars, \core\ \fireTwo\ galaxies as diamonds and \tripleLatte\ as a square, and the observations from the left panel are repeated in grey for reference. For the observed galaxies we plot \twelvePlusOH\ as reported in the xCOLD GASS and KLM papers, which is inferred from ratios of collisionally excited emission lines. For the \fireTwo\ galaxies, we instead calculate \twelvePlusOH\ directly from the oxygen and hydrogen masses of the gas particles within the half-light radius, \RvisibleHl, i.e., we do not model the emission lines of the gas and infer the metallicity from line ratios as is done for the observations. For the \fireTwo\ galaxies we are plotting using \sfrHundredMyr. The lines show the median \LcoOneZero\ in bins of SFR for all \fireTwo\ galaxies combined, for the xCOLD GASS detections and for KLM.

The xCOLD GASS detections follow a nearly linear relation between \LcoOneZero\ and SFR, with $\rm \LcoOneZero\propto SFR^{0.8}$ and a relatively small scatter of $\approx0.4$ dex at all SFRs. The KLM compilation is considerably more sensitive and reaches luminosities as low as $\rm \LcoOneZero\approx10^5\,\obsunits$, with 10 of its 53 galaxies fainter than the faintest xCOLD GASS detection. Analogous sources appear only as upper limits in xCOLD GASS because of its relatively high detection limit, which ranges from $\rm \LcoOneZero\approx 4\times10^8\,\obsunits$ to $3\times10^7\,\obsunits$ depending on stellar mass. As a result, xCOLD GASS detects CO in only $\approx40\%$ of the galaxies with $\rm SFR > 0.3\,M_\odot\,yr^{-1}$. Including these faint emitters changes the relation between \LcoOneZero\ and SFR substantially. The KLM data yield a slope twice as steep, $\rm \LcoOneZero\propto SFR^{1.6}$, and a much larger scatter of $\approx0.7$--$1$ dex.

The same contrast between the two samples appears in the relation between \LcoOneZero\ and \twelvePlusOH. At a fixed SFR of $1\,\rm M_\odot\,yr^{-1}$, a 0.4 dex increase in oxygen abundance corresponds to an increase of $\approx0.6$ dex in \LcoOneZero\ for xCOLD GASS, giving a slope of ${\rm d}\log\LcoOneZero/{\rm d}(\twelvePlusOH)\approx1.5$. The KLM compilation, which includes fainter and more metal poor CO emitters, yields a steeper slope of $\approx2.4$. In both datasets, the relation flattens above $\twelvePlusOH \sim 8.6$, where the slope drops below unity.

Now we turn our attention to our simulation predictions. All \fireTwo\ suites follow trends similar to those of \firebox, our lowest resolution run. For $\rm \sfrHundredMyr > 3\,M_\odot\,yr^{-1}$, the small sample of \core\ \fireTwo\ galaxies yields \LcoOneZero\ values about 0.3 dex higher than the \firebox\ median at fixed SFR, while the \tripleLatte\ run, with $8\times$ better mass resolution, lies only slightly above the median relation. However, they are always within the $1\sigma$ scatter of the \firebox\ distribution, therefore, we find no statistically significant dependence of our \LcoOneZero\ predictions on resolution over the range probed here.

Compared with the observations, the \fireTwo\ predictions agree well with both xCOLD GASS data and KLM compilation for $\rm 3 \leq SFR/(M_\odot\,yr^{-1}) \leq 30$. Our predictions lie factor of three below the xCOLD GASS data, yet remain within the scatter of the more sensitive KLM compilation. However, for $\rm SFR \leq 1\,M_\odot\,yr^{-1}$. our \fireTwo\ galaxies yield increasingly bigger discrepancy between xCOLD GASS. However, our \fireTwo\ predictions are always within the scatter of the more sensitive KLM compilation. 

A similar pattern emerges in the scatter of \LcoOneZero\ at fixed SFR. For $\rm SFR \gtrsim 3\,M_\odot\,yr^{-1}$ where xCOLD GASS detects CO in all of the observed galaxies, \fireTwo\ predicts a scatter of $\approx0.3$ dex identical to that of xCOLD GASS. Toward lower SFRs the \fireTwo\ scatter increases considerably, reaching 0.5 dex for $\rm 0.5 \leq SFR/(M_\odot\,yr^{-1}) \leq 3$ and 0.8 dex for $\rm SFR \lesssim 0.5\,M_\odot\,yr^{-1}$, while over the same range xCOLD GASS scatter stays same. Meanwhile the KLM compilation shows 0.7 dex and 1.0 dex in the same SFRs, in close agreement with our predictions. 

Taken together, these comparisons suggest that \fireTwo\ reproduces the observed scatter wherever the detection rates is high and the observations reliably sample the full distribution. Toward lower SFRs, an increasing fraction of faint CO emitters appear only as upper limits in xCOLD GASS. Excluding them narrows the measured scatter and biases the median upward, which would explain why our predictions diverge from xCOLD GASS in this regime while remaining consistent with the KLM compilation. More sensitive CO surveys at low SFR would provide a direct test of the scatter predicted by \fireTwo.

We now turn to the dependence on metallicity. At fixed SFR, our intrinsic \twelvePlusOH\ values are higher than the metallicities inferred for both xCOLD GASS and KLM. The offset grows with SFR, from about $\rm 0.1\,dex$ at $\rm SFR \approx 0.1$--$\rm 0.3\,M_\odot\,yr^{-1}$ to $\rm 0.5\,dex$ at $\rm SFR \approx 1$--$\rm 3\,M_\odot\,yr^{-1}$, and exceeds $\rm 0.7\,dex$ for $\rm SFR > 3\,M_\odot\,yr^{-1}$. This offset most likely reflects the difference between the intrinsic metallicities measured directly in the simulations and those inferred observationally from collisionally excited lines. Different calibrations of these lines can disagree by $\rm 0.5$--$\rm 0.8\,dex$ \citep{kewleyMetallicityCalibrationsMassMetallicity2008, Moustakas_2010_krumholzCompiledData}, so metallicities are best compared using a single, consistent calibration. Because we do not have the tools to do this for this work, we instead we focus on the slope of the relation between \LcoOneZero\ and metallicity rather than on a direct comparison of \LcoOneZero\ at fixed SFR and metallicity. Our \fireTwo\ predictions yield a slope of ${\rm d}\log\LcoOneZero/{\rm d}(\twelvePlusOH)\approx2.4$, nearly identical to that of the KLM compilation at $\rm SFR \approx 1\,M_\odot\,yr^{-1}$ and steeper than the $\approx1.5$ found for xCOLD GASS.

Overall, our \fireTwo\ predictions reproduce both the mean \LcoOneZero--SFR relation and its scatter, particularly when compared with the more sensitive KLM compilation. Our predicted \LcoOneZero--\twelvePlusOH\ relation also has a slope close to that of the KLM compilation. There is, however, some indication that \fireTwo\ underpredicts \LcoOneZero\ at fixed metallicity, although this comparison is limited by the large systematic uncertainties and possible mismatch between inferred metallicities, which is a known problem.

\subsection{\LcoOneZero\ luminosity function at $z=0$}

\begin{figure}[ht!]
    \centering
    \includegraphics[width=0.99\linewidth]{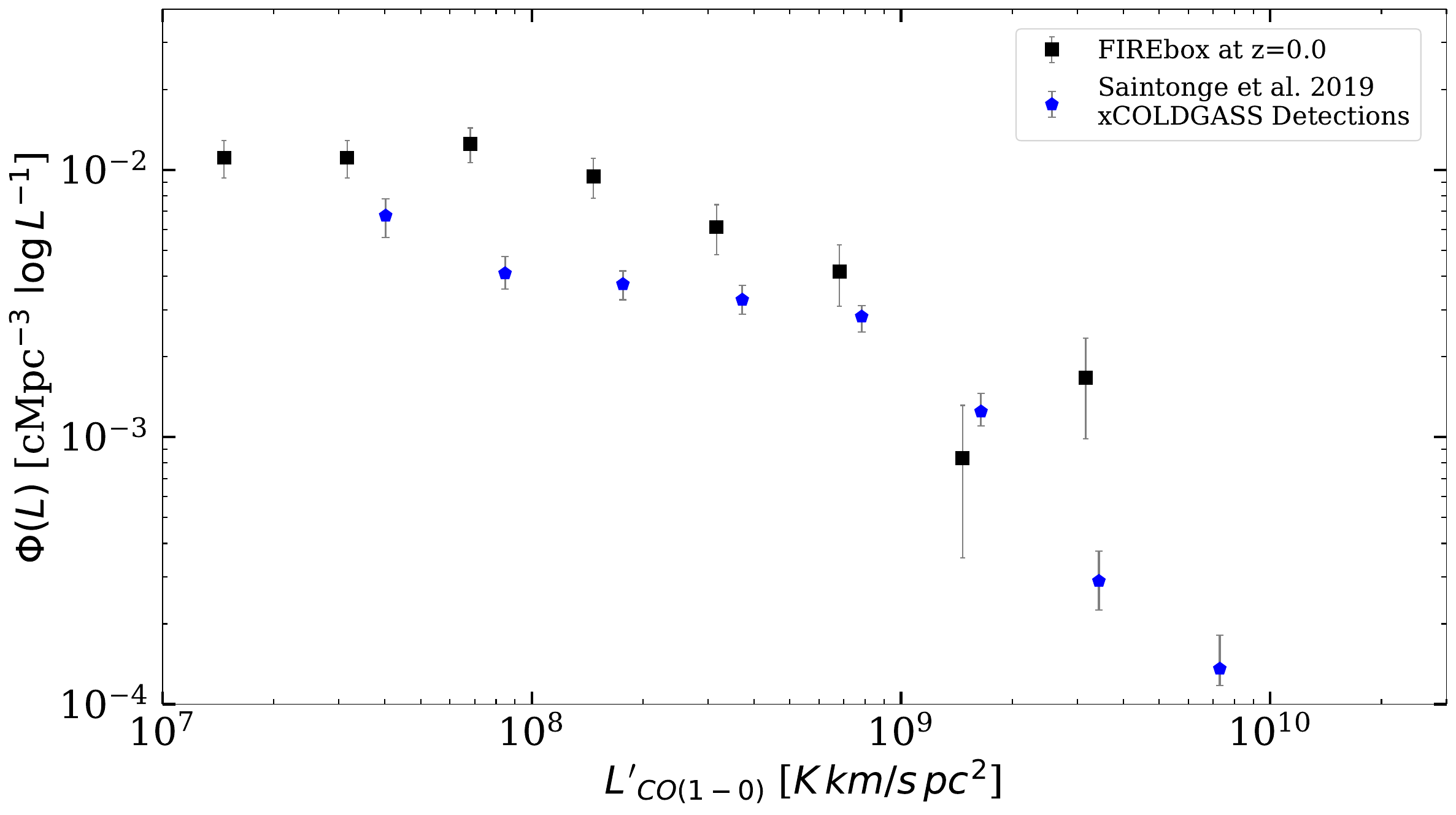}
    \caption{The CO(1-0) luminosity function calculated from \firebox\ simulations at $z=0$ (black squares), and that determined from the xCOLD GASS survey \citep{Saintonge_2017} (filled blue circles, with their best fit Schechter function shown as the blue dotted line). The xCOLD GASS measurements are complete down to $\LcoOneZero = 10^{7.5}\obsunits$, so we show the observational data only above this limit. The \firebox\ galaxies are divided into 10 logarithmic bins over the range $7 < \log(\LcoOneZero) < 10$. The error bars for each \firebox\ bin account for Poisson statistics only, calculated using Equation~\ref{eq:possion_uncertainty}. For $10^8<\LcoOneZero/(\obsunits)<10^9$, the \firebox\ estimates are within roughly one standard deviation of the xCOLD GASS results.}
    \label{fig:co10_lf_z0_observations}
\end{figure}

Figure~\ref{fig:co10_lf_z0_observations} compares the CO(1-0) luminosity function (LF) of \firebox\ galaxies more massive than $\rm \Mstar = 10^9\,M_\odot$ against the xCOLD GASS detections. We show \firebox\ predictions with the black squares and xCOLD GASS data with blue circles. We apply a stellar mass cut to the \firebox\ LF so that it matches the mass selection of the xCOLD GASS survey. We calculated the \firebox\ CO(1-0) LF by counting galaxies in logarithmic luminosity bins and normalizing by the simulation volume and bin width by:
\begin{equation}
    \Phi_{\LcoOneZero} = \frac{N}{V \, \Delta\log(\LcoOneZero)} 
\end{equation}
where $N$ is the number of central \firebox\ galaxies with $\rm \Mstar > 10^9\,M_\odot$ in each bin, $\rm V=(22.1\,cMpc)^3$ is the comoving volume of the simulation, and $\rm\Delta\log(\LcoOneZero)=0.3\,dex$ is the bin width.

The error bars on the \firebox\ predictions account for Poisson statistics only (not accounting for variations in SFR, $Z$, etc.). We calculate the Poisson uncertainty as
\begin{equation}
    \label{eq:possion_uncertainty}
    \epsilon_{\Phi_{\LcoOneZero}} = \frac{\sqrt{N}}{V \, \Delta\log(\LcoOneZero)} ,
\end{equation}
The number density of \firebox\ CO emitters exceeds the observed one in nearly every luminosity bin. The tension increases at around $\LcoOneZero\approx2\times10^8\obsunits$, where it exceeds $4\sigma$. This luminosity is close to the xCOLD GASS sensitivity limit for its most massive galaxies, so the deficit of observed CO emitters in this regime likely reflects incompleteness in the data rather than a genuine discrepancy.

At higher luminosities, the excess drops to the $1$--$3\sigma$ level, and part of this remaining offset may arise from cosmic variance, since \firebox\ is a moderately sized volume. Figure A1 of \cite{Feldmann_2023} shows that the unweighted halo mass function (HMF) of the \firebox\ volume can exceed the fiducial HMF by up to $\rm\approx0.2\,dex$ at the highest halo masses, which would inflate the number of bright CO emitters in our LF. \cite{Feldmann_2023} show that this discrepancy disappears once the uniform $1/V$ weights are replaced with an appropriate weighting scheme. Since we apply no weighting when computing our LFs, adopting their reweighting could bring our predictions into better agreement with the data at the bright end.

\subsection{\LcoTwoOne\ versus SFR}
\subsubsection{\LcoTwoOne\ versus SFR at $z=0$}

\begin{figure}[h!]% Figure 6 L_CO 2-1 vs SFR
    \centering
    \includegraphics[width=0.99\linewidth]{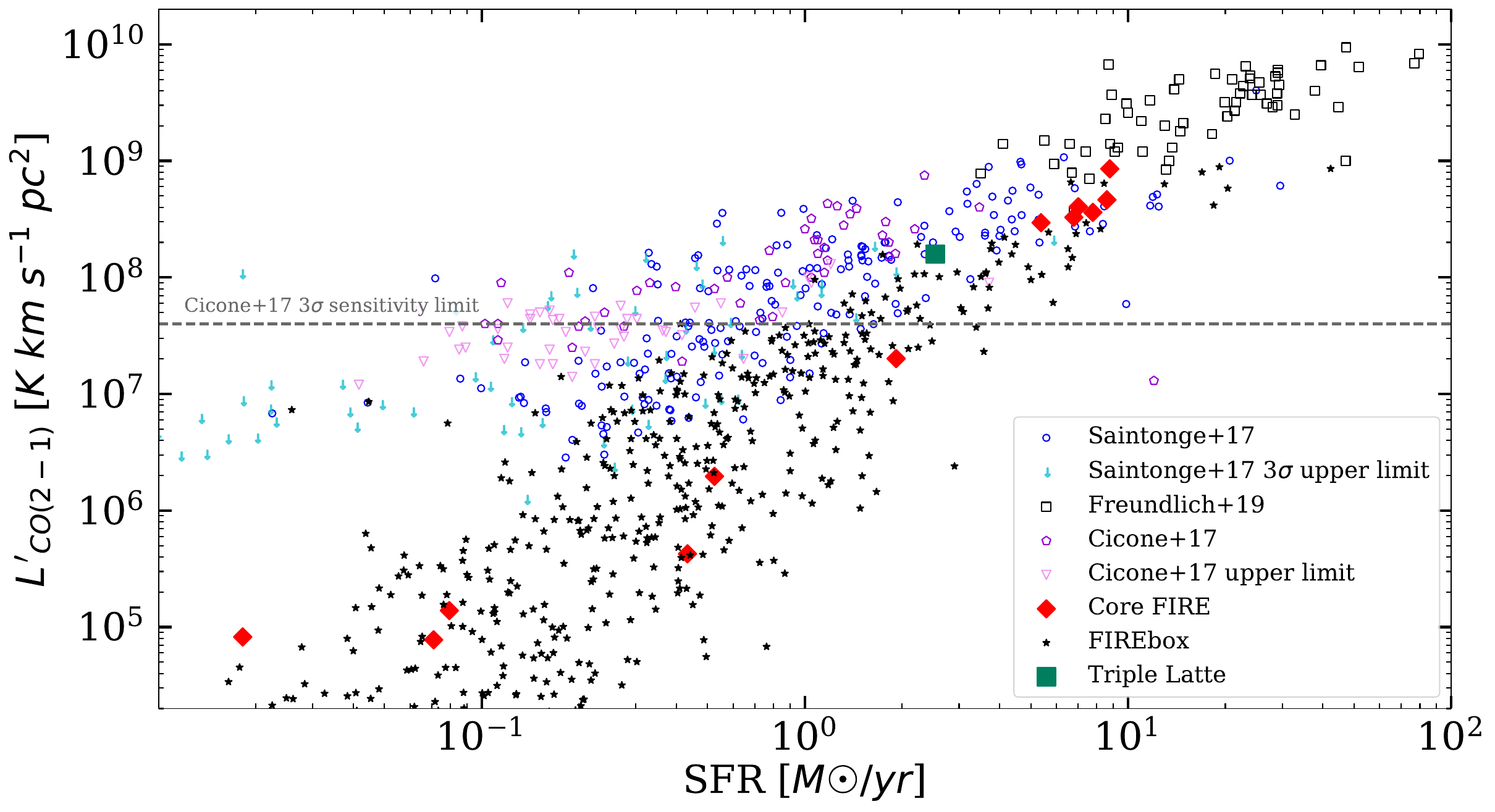}
    \caption{The predicted CO(2–1) luminosity as a function of SFR for the post-processed FIRE simulations (\firebox\ with black stars, \core\ \fireTwo\ with red diamonds, and Triple Latte with green squares). Also shown are measurements from the xCOLD GASS \citep{Saintonge_2017} ($0.01 < z < 0.05$; open blue circles, with downward arrows indicating upper limits), PHIBSS2 \citep{Freundlich_2019} ($0.5 < z < 0.8$; open black squares), and ALLSMOG \citep{Cicone_2017_allsmog} ($0.01 < z < 0.03$; purple pentagons, with downward triangles indicating upper limits) surveys. The dashed gray line marks the sensitivity limit of the ALLSMOG which is at $4\times10^7\,\obsunits$. The predicted CO(2-1) luminosities of FIRE galaxies with SFR $\gtrsim 7\,M_\odot\,{\rm yr}^{-1}$ are consistent with the observations, while at the same time the \fireTwo\ mean is nearly a factor of two lower than the observations while the predictions at $SFR\leq 2$ show substantial scatter This may reflect the limited sensitivity of the surveys, see Table~\ref{table:observational_data}}
    \label{fig:Lco21_vs_sfr10Myr_galaxyType_z0}
\end{figure}

Figure \ref{fig:Lco21_vs_sfr10Myr_galaxyType_z0} plots the CO(2-1) emission at z=0 as a function of SFR for xCOLD GASS, PHIBSS2 \citep{Freundlich_2019} and ALLSMOG \citep{Cicone_2017_allsmog} detections and upper limits, as well as the results of our \fireTwo-based calculations. For $SFR \geq7\,M_\odot\,\text{yr}^{-1}$ the mean $\mu(\log(L'_{CO(2-1)}))$ is slightly lower than observed. At the same SFR range the \fireTwo\ predictions are producing lower scatter than the observations.

At $\rm SFR \lesssim 2\,M_\odot\,{\rm yr}^{-1}$, the FIRE-predicted $L'_{CO(2-1)}$ increase slightly more rapidly with SFR than the observations appear to. The \fireTwo\ galaxies also exhibit much larger scatter, and reach to much lower values, of $L'_{CO(2-1)}$ than the observations at the same SFR. On average, the low SFR \fireTwo\ sample is nearly an order of magnitude fainter than the xCOLD GASS detections and nearly two orders of magnitude fainter than the ALLSMOG detections. We note that the sensitivity limit of the ALLSMOG is $\LcoTwoOne=4\times10^7\,\obsunits$ and there are no observational detections at luminosities below $\LcoTwoOne\approx10^6\,\obsunits$ in the surveys we plot. This a factor of ten higher than the luminosities seen in the KLM compilation. The ratio of \LcoTwoOne\ to \LcoOneZero\ is typically $0.6$ or so \citep{2021MNRAS.504.3221D, 2024ARA&A..62..369S}. This strongly suggests that more sensitive surveys will be needed to test the \fireTwo\ \LcoTwoOne\ predictions.

\subsubsection{\LcoTwoOne\ versus SFR at $z \sim 0.5-1.4$}

%-------------------------------------------------------------------------
\begin{figure}[h!]
    \centering
    \includegraphics[width=0.99\linewidth]{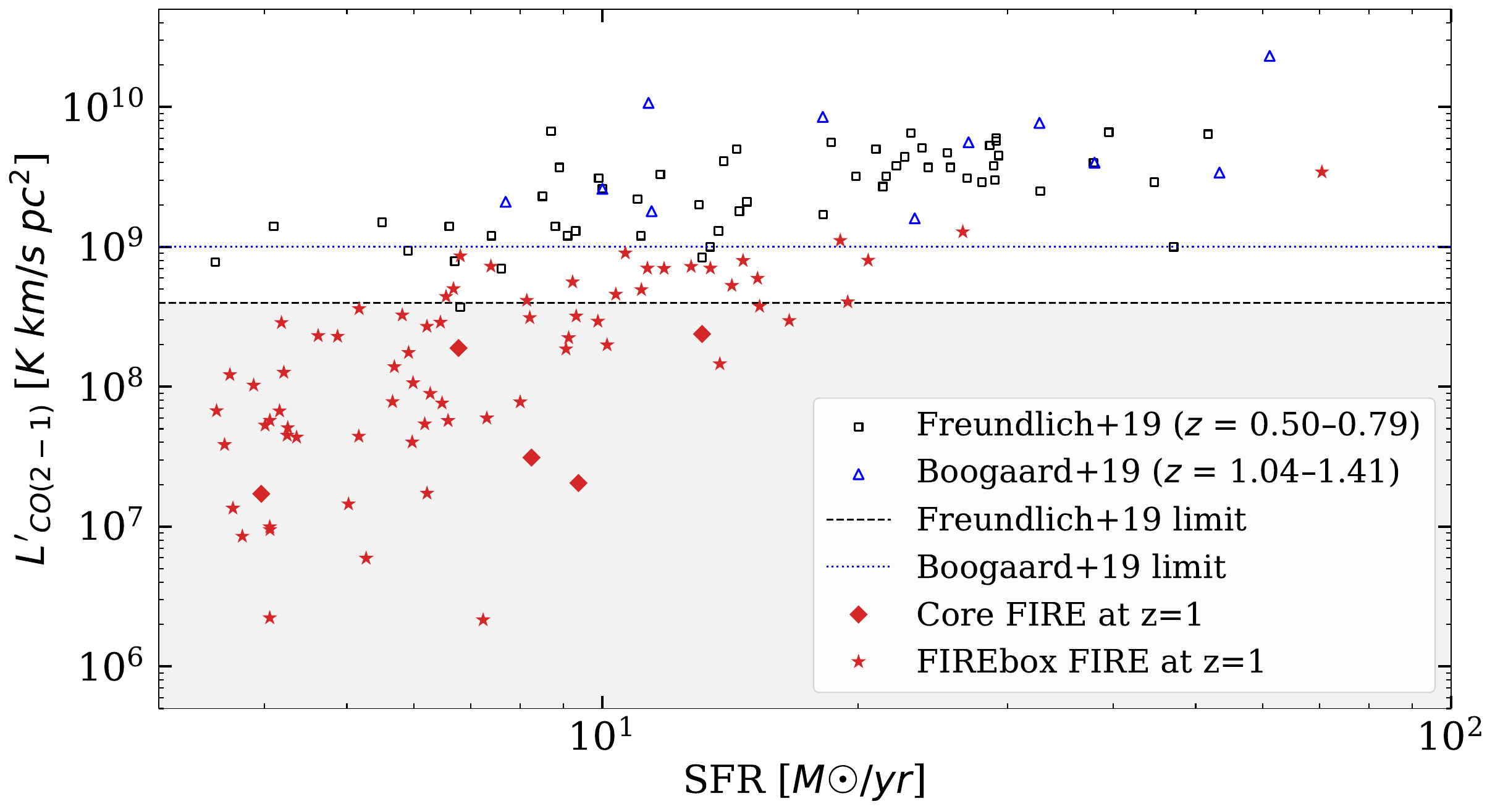}
    \caption{\LcoTwoOne\ as a function of SFR for $0.5\lesssim z\lesssim1.4$.  PHIBSS2 observations \citep{Freundlich_2019}, spanning $0.50 < z < 0.79$, are shown as unfilled black squares, and ASPECS observations  at redshifts $1.04 < z < 1.41$ \citep{boogaard_2019_alma_aspecs} as unfilled blue triangles. We denoted the lowest \LcoTwoOne\ detected by PHIBBS2 with dashed line and the sensitivity limit of the ASPECS with dotted line. \core\ \fireTwo\ and \firebox\ simulations with $\Mstar/M_\odot \geq 10^{10}$ at z=1 are represented by diamonds and stars, respectively. }
    \label{fig:Lco21_vs_sfr100Myr_galaxyType_z1_matchedMstarSFR}
\end{figure}
%-------------------------------------------------------------------------

In this section we compare the CO(2–1) emission of our post-processed \fireTwo\ galaxies with moderate-redshift observations from the PHIBSS2 ($0.5<z<0.8$) \citep{Freundlich_2019} and ASPECS ($1<z<3$) \citep{boogaard_2019_alma_aspecs} surveys, which we describe in Section~\ref{sec:observational_data}. These are the redshift ranges of the full samples, but we only use the galaxies that can be directly compared with our $\rm z=1$ snapshot. For ASPECS, we only take the galaxies detected in CO(2–1). The CO(2-1) detected galaxies in the survey lie at $1.04<z<1.55$. The higher redshift ASPECS galaxies are only detected in CO(3–2) or CO(4–3). Comparing them with our CO(2–1) predictions would require us to assume excitation ratios between the CO lines, so we leave them out. 

All PHIBSS2 galaxies are observed in CO(2–1) at $0.50<z<0.79$. The redshifts of the detected galaxies lies slightly below our \fireTwo\ galaxies presented in this section. To further limit both the \fireTwo\ galaxies and the observations to the stellar mass ($5.0\times10^9 < \Mstar / M_\odot < 2.6\times10^{11}$) and SFR ($\rm 3.5 < SFR/(M_\odot\,yr^{-1}) < 70$) ranges shared by all three samples. After this cut, the PHIBSS2 galaxies span $0.50<z<0.79$ and the ASPECS galaxies span $1.04<z<1.41$. We show the CO(2–1) luminosity of these galaxies as a function of SFR in Figure~\ref{fig:Lco21_vs_sfr100Myr_galaxyType_z1_matchedMstarSFR}.

The observations follow a fairly flat relation, with \LcoTwoOne\ increasing by only a factor of $\approx5$ between $\rm SFR = 4$ and $\rm 60\,M_\odot\,yr^{-1}$, corresponding to a slope of $\rm d\log(\LcoTwoOne)/d\log(SFR)\approx 0.6$. Our simulations, in contrast, yield a much steeper relation with a slope of $\approx1.4$. At fixed SFR, the \fireTwo\ galaxies. The median \fireTwo\ prediction lies $\approx1$--$1.5$ dex below the data for $\rm 4 < SFR/(M_\odot\,yr^{-1}) < 10$, but because of the steeper slope, the offset decreases to $\approx0.7$--$1$ dex for $\rm SFR > 10\,M_\odot\,yr^{-1}$.

At $\rm SFR \approx 4$ to $\rm 6\,M_\odot\,yr^{-1}$, the \LcoTwoOne\ values predicted by our \fireTwo\ galaxies range from about $2\times10^6$ to $4\times10^{8}\,\obsunits$, spanning more than two orders of magnitude, while the observations show far less scatter. This difference most likely reflects detection limits. Because the observed samples are effectively restricted to $\LcoTwoOne \geq 4\times10^8\,\obsunits$, the sharp lower boundary of the observed distribution is probably a consequence of survey sensitivity rather than a physical floor. What this figure demonstrates, then, is simply that the \fireTwo\ sample includes galaxies too faint for these surveys to detect.

The observed galaxies also have much higher gas fractions and slightly longer depletion times than most of our \fireTwo\ sample. In Appendix~\ref{sec:Lco21_z1_fgas_tdep_Mstar_sfr_matched} we match galaxies in stellar mass, SFR, gas fraction ($\rm f_{gas} = M_{mol}/(M_{mol} + \Mstar)$), and depletion time. For this matched sample, the \fireTwo\ galaxies lie only about a factor of five below the observations, although with nearly twice their scatter. This larger scatter is driven by the wide range of metallicities among the \fireTwo\ galaxies.

\subsection{Evolution of \LcoOneZero\ with Redshift}

\begin{figure}[h!]
    \centering
    \includegraphics[width=0.99\linewidth]{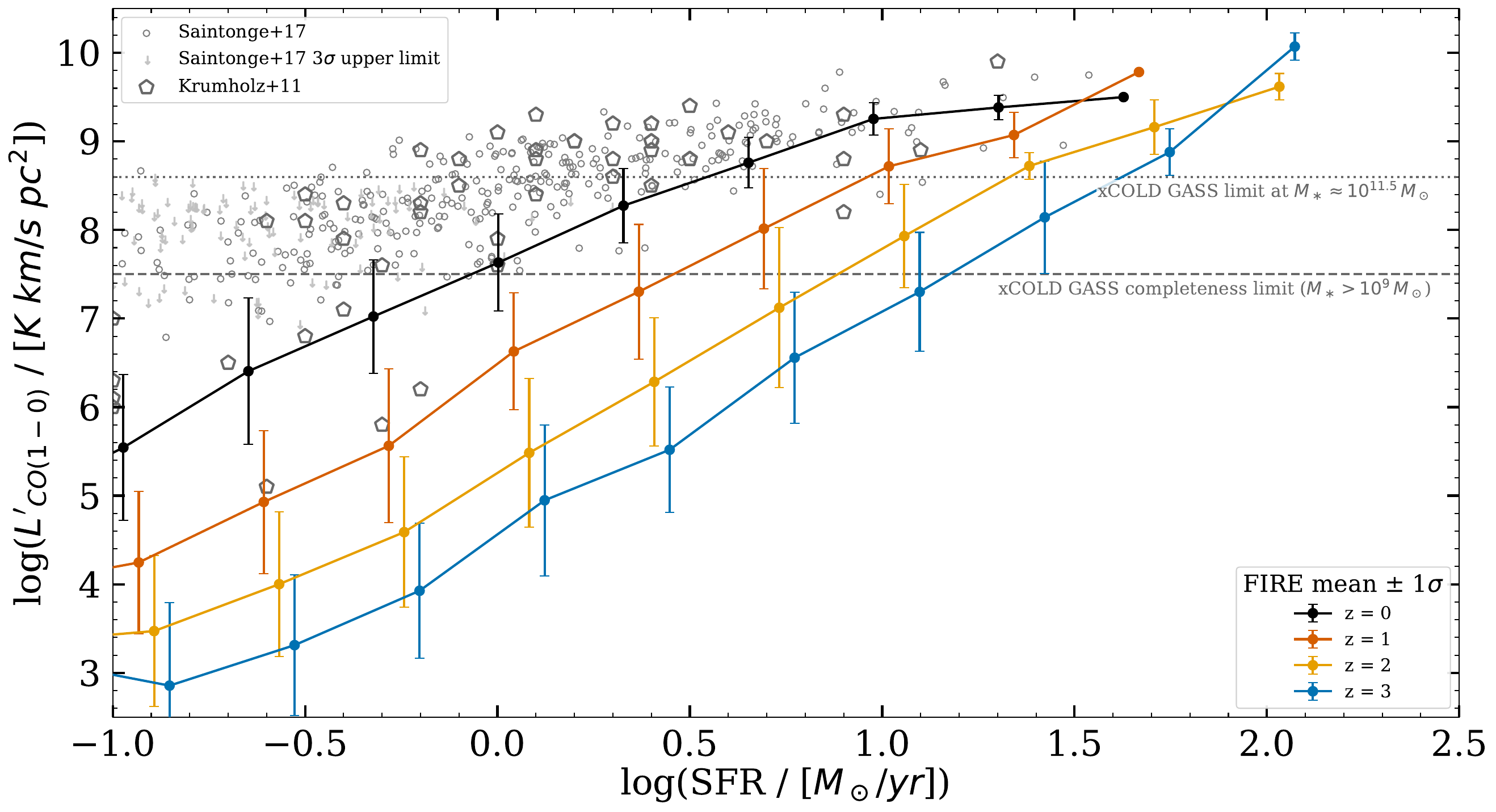}
    \caption{FIRE-based predictions of \LcoOneZero\ as a function of SFR are displayed for \(z=0\) (black), \(z=1\) (red), \(z=2\) (orange), and \(z=3\) (blue). The \fireTwo\ galaxies are divided into 20 bins over the interval \(-4 < \log(\mathrm{SFR}) < 2.5\). Within each bin, the solid line indicates the mean logarithmic \LcoOneZero\ of the \fireTwo\ galaxies, and the error bars marks the \(1\sigma\) scatter around this mean. For clarity, the points at different redshifts are offset horizontally by up to 0.06\,dex. For comparison, we include in gray the \(z=0\) observations from \cite{Saintonge_2017} (open circles for detections and arrows for \(3\sigma\) upper limits), together with the compiled galaxy sample from \cite{Krumholz_2011_lowZ_z0_observations} (open pentagons), as presented in Figure \ref{fig:Lco10_vs_sfr10Myr_galaxyType_z0_colorBarLogOH}. The dashed gray line marks \(\log(\LcoOneZero)=7.5\), below which the xCOLD GASS survey \citep{Saintonge_2017} becomes incomplete due to its stellar mass cut (\(M_\ast > 10^{9}\,M_\odot\)), while its gas fraction limited CO sensitivity varies with stellar mass and reaches only \(\log(\LcoOneZero)\approx 8.6\) for the most massive galaxies. At fixed SFR, our results predict that CO luminosity drops markedly toward higher redshift, although galaxies with higher SFRs rapidly converge onto the $\rm z=0$ distribution.}
    \label{fig:Lco_10_z0_z1_z2_z3_sfr100Myr}
\end{figure}

Here we compare the predictions of our \fireTwo\ galaxies at four different redshifts, from z=0 to z=3. Figure \ref{fig:Lco_10_z0_z1_z2_z3_sfr100Myr} shows \LcoOneZero\ as a function of SFR calculated for \fireTwo\ galaxies at these four different redshifts. We replot the $z=0$ $L'_{CO(1-0)}$ observations of xCOLD GASS and KLM for reference. 

At low SFRs, the relation has a similar power law slope at all redshifts. Meanwhile, the normalization of the relation varies considerably with redshift. For $\rm \sfrHundredMyr < 6\,M_\odot\,yr^{-1}$, our \fireTwo\ galaxies yield a superlinear linear, $\rm \LcoOneZero \propto SFR^{n}_{100Myr}$, with $n\approx2.1$--$2.4$ at all redshifts, somewhat steeper than the slope of $\approx1.6$ found for the KLM compilation. At fixed SFR, however, the mean \LcoOneZero\ of our \fireTwo\ predictions decreases nearly an order of magnitude at each increasing redshift. The change in the normalization gets weaker with increasing redshift and about 0.5-0.7 dex between z=2 to z=3.

Meanwhile, at the higher SFRs, the relation between \LcoOneZero\ and \sfrHundredMyr\ starts differ with redshift. At $\rm z=0$, the relation flattens around $\rm \sfrHundredMyr \approx 6\,M_\odot\,yr^{-1}$, which is also consistent with the KLM and xCOLD GASS data at the same redshift. We also see a hints of flattening at z=1 and z=2; albeit at higher SFRs. The SFR at which the flattening occurs increases with redshift, and at z=3 we see no flattening up to the highest SFRs we can probe ($\rm \sfrHundredMyr \approx 200\,M_\odot\,yr^{-1}$), which are occupied by the \massivefire\ halos. Because the low redshift relation flatten while the high redshift ones remain steep, the gap between mean \LcoOneZero\ at different redshifts rapidly narrows with increasing SFR.

The scatter in our \fireTwo\ predictions show a weak dependence on redshift. For $\rm 0.1 < \sfrHundredMyr/(M_\odot\,yr^{-1}) < 6$, we find $\sigma_{\log(\LcoOneZero)}\approx0.8$ dex at $z>0$ and a slightly smaller $\approx0.6$ dex at $z=0$. At higher SFRs, the scatter decreases at all redshifts. The decrease in the scatter can be attributed to the small number of simulated galaxies in this regime.

\subsubsection{Dependence of $\LcoOneZero$ on metallicity, stellar mass, and redshift \label{sec:Lco10_dependence_on_Z_Mstar_z}}

\begin{figure*}
    \centering
    \includegraphics[width=0.99\linewidth]{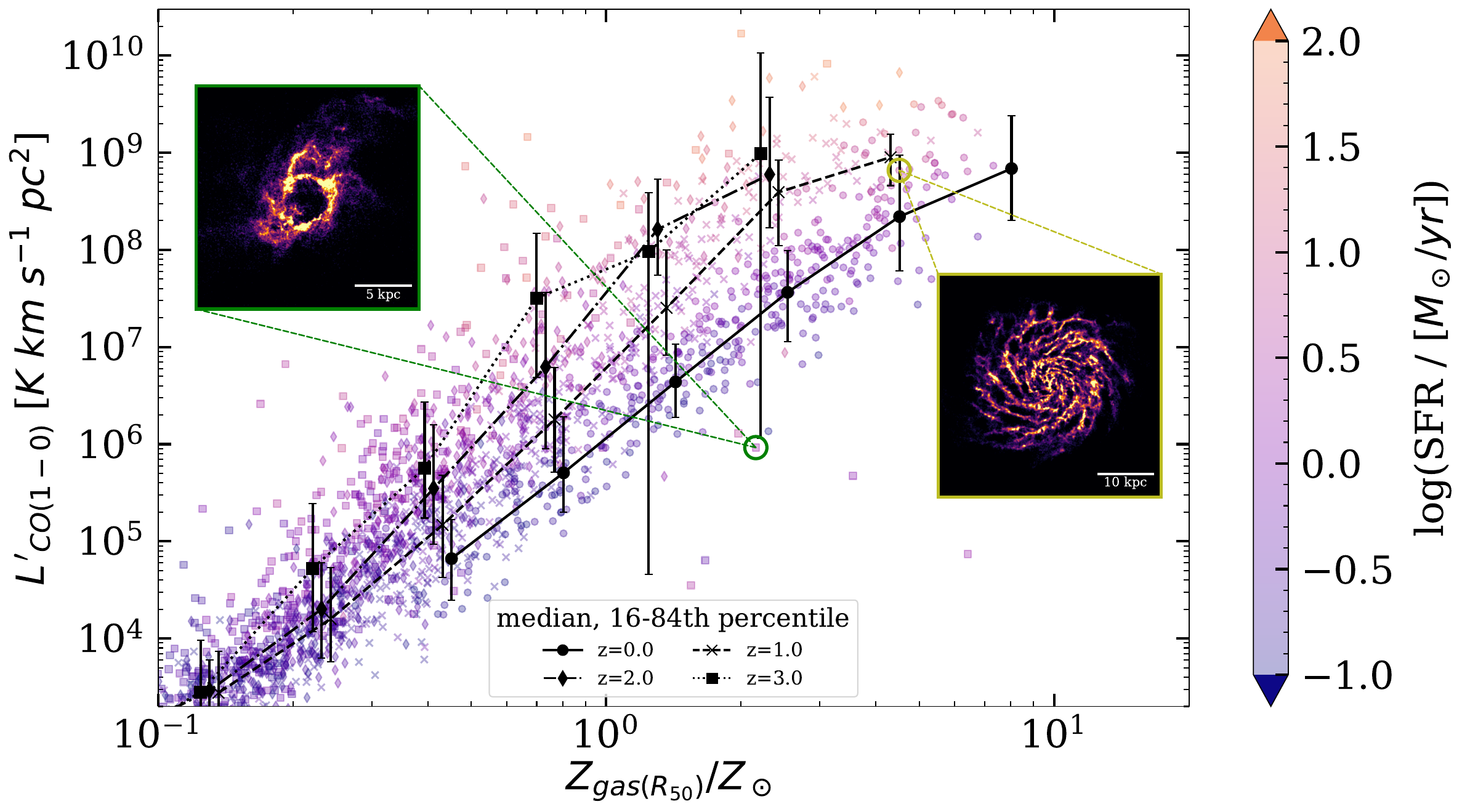}
    \caption{\LcoOneZero\ of \fireTwo\ galaxies with $\rm \sfrHundredMyr > 0.1\,M_\odot\,yr^{-1}$ as a function of the mass-averaged gas metallicity within the half-light radius (\ZgasHl), normalized to the solar value of 0.0134 \citep{Grevesse2010_solarMetallicity}. Galaxies at \(z=0\), 1, 2, and 3 are shown as circles, crosses, diamonds, and squares, respectively, and are colored by their SFR averaged over the past 100\,Myr. The black symbols connected by lines (solid, dashed, dash-dotted, and dotted for \(z=0\), 1, 2, and 3) show the median \LcoOneZero\ in bins of 0.25\,dex in \ZgasHl\ at each redshift, and the error bars span the 16th to 84th percentiles; only bins with at least five galaxies are shown, and the medians at different redshifts are slightly offset horizontally for clarity. \LcoOneZero\ increases steeply with metallicity at all redshifts, and at fixed metallicity galaxies with higher SFR are brighter in CO. The green circle marks one central \firebox\ halo at $z=3$, whose gas surface density map is shown in the upper-left inset (the scale bar is 5\,kpc). Its gas is distributed in a ring around a central cavity probably due to recent starburst which expelled most of the gas from the center of the galaxy. For comparison, the olive circle marks another central \firebox\ galaxy at $z=0$ that lies on the $z=0$ median relation; its gas surface density map is shown in the right inset (the scale bar is 10\,kpc) and shows a gas-rich spiral disc without a central cavity.
}
    \label{fig:Lco10_z0_z1_z2_z3_metallicity}
\end{figure*}

\begin{figure*}
    \centering
    \includegraphics[width=0.99\linewidth]{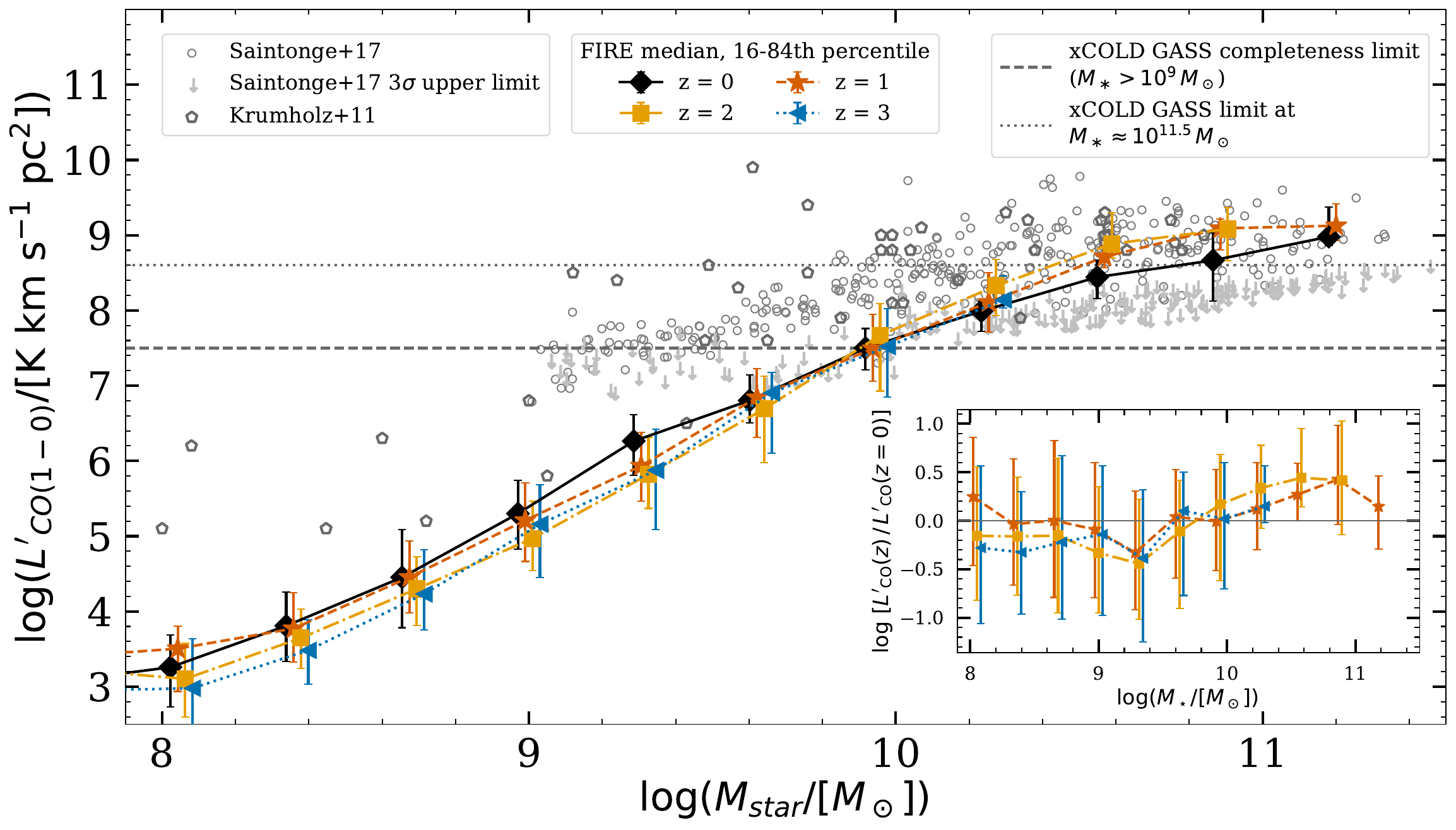}
    \caption{\LcoOneZero\ as a function of galaxy stellar mass at different redshifts. For the \fireTwo\ galaxies, \Mstar\ is the total mass of all star particles within $20\,\kpc$ of the halo center (and within $|z|<5\,\kpc$ of the disc plane for galaxies with a well-defined disc). The local galaxies from xCOLD GASS are shown in gray as open circles for detections and downward-pointing arrows for the \(3\sigma\) upper limits, and the galaxies from KLM as dark gray pentagons. xCOLD GASS observed each galaxy until it was detected or until a limiting molecular gas fraction is reached. Its \LcoOneZero\ sensitivity therefore depends on stellar mass and it is $\log(\LcoOneZero)\approx7.5$ for the lowest-mass galaxies ($\Mstar \approx 10^{9}\,M_\odot$, dashed gray line) and rises to $\log(\LcoOneZero)\approx8.6$ for the most massive galaxies ($\Mstar \approx 10^{11.5}\,M_\odot$, dotted gray line). The \fireTwo\ galaxies are divided into 19 equal-width bins over \(6 < \log(\Mstar/M_\odot) < 12\). The black diamonds, red stars, orange squares, and blue triangles, connected by solid, dashed, dash-dotted, and dotted lines, show the median $\log(\LcoOneZero)$ in each bin at \(z=0\), 1, 2, and 3, respectively, and the error bars span the 16th to 84th percentiles. Only bins with at least five galaxies are shown, and the symbols at different redshifts are slightly offset horizontally for clarity. The inset shows the difference between the medians at each redshift and at $z=0$, $\log[L'_{\rm CO}(z)/L'_{\rm CO}(z=0)]$, with error bars combining in quadrature the 16th to 84th percentile ranges of $\log\LcoOneZero$ at the two redshifts. At fixed \Mstar, the medians at all redshifts agree to within $\approx0.5$\,dex, a much weaker redshift dependence than at fixed SFR (Figure~\ref{fig:Lco_10_z0_z1_z2_z3_sfr100Myr}). \label{fig:Lco_vs_Mstar_redshifts}}
\end{figure*}

Our calculations predict a dramatic decrease in \LcoOneZero\ at a given SFR as redshift increases. Why should this be the case? Here we show that the trend is explained in part by the variation of metallicity and star formation rate with redshift.

We start by plotting the predicted \LcoOneZero\ of \fireTwo\ galaxies at redshifts $z = 0, 1, 2,$ and $3$ as a function of mass-averaged gas metallicity for all gas particles inside the half-light radius (\ZgasHl) and SFR within 100 Myrs (\sfrHundredMyr) (depicted by color) in Figure~\ref{fig:Lco10_z0_z1_z2_z3_metallicity}. In the figure we only show the galaxies with $\rm \sfrHundredMyr > 0.1\,M_\odot\,yr^{-1}$. At the same metallicity both the SFR and \LcoOneZero\ of \fireTwo\ galaxies increase with redshift. For about two orders of magnitude change in the \ZgasHl, the \LcoOneZero\ changes by about six orders of magnitude, Meanwhile for supersolar galaxies the relation is somewhat shallower. This is expected because once CO is fully shielded, additional shielding produces little extra \LcoOneZero. Instead, a higher CO column density mainly increases the optical depth of the gas cells, so \LcoOneZero\ starts to saturate.

Careful examination of the figure reveals that \fireTwo\ galaxies at different redshifts but with similar metallicity and SFR tend to produce comparable \LcoOneZero, although even then there is substantial scatter. The fact that similar metallicities are associated with similar \LcoOneZero\ shows that the offset between redshifts in Figure~\ref{fig:Lco_10_z0_z1_z2_z3_sfr100Myr} at a fixed SFR is driven in part by metallicity differences, but the residual scatter suggests that there is another variable at play. 

The bursty star formation could partly drive the scatter in \fireTwo\ predictions. Galaxies with shallow gravitational potentials cannot hold on to their gas against stellar feedback,
which makes their star formation highly bursty. In Figure~\ref{fig:Lco10_z0_z1_z2_z3_metallicity} we plot one of the central halo at $\rm z=3$ in the \firebox\ volume that probably underwent a recent starburst activity. The intense stellar feedback removed much of the gas from the galaxy's central region, producing a central cavity with a surrounding ring-like gas distribution. The remaining gas inside the half light radius is still highly enriched, i.e. $\rm \ZgasHl\approx2\,Z_\odot$, yet the molecular gas is disrupted, drastically reducing the CO luminosity retained from this halo.

Stellar mass, which tracks both metallicity and SFR, therefore may be the more fundamental parameter governing \LcoOneZero. To test this, Figure~\ref{fig:Lco_vs_Mstar_redshifts} shows \LcoOneZero\ as a function of \Mstar\ for our \fireTwo\ galaxies at four redshifts.

In the inset of the same figure we show the relative difference between the \fireTwo\ predictions at higher redshifts against the $\rm z=0$ distribution. For each stellar mass bin, we took the median \LcoOneZero\ of all galaxies in that bin at a given redshift and subtracted the median of the same bin at $z=0$. This difference, $\Delta \log L'_{\rm CO}$, is what the inset shows. The error bars are showing the distance from the median to the 16th and 84th percentiles for each redshift. A positive value means that galaxies at redshift z are brighter in CO(1-0) than galaxies of the same stellar mass at $\rm z=0$

In \fireTwo, the relation between \LcoOneZero\ and \Mstar\ shows almost no dependence on redshift. At fixed stellar mass, the \LcoOneZero\ values at different redshifts differ by at most a factor of three. Nevertheless the difference is always smaller than the scatter at individual redshifts. 

The relative insensitivity of \LcoOneZero\ to redshift at a given \Mstar\ (as opposed to the strong sensitivity to $z$ at a given SFR) is not unexpected. This is apparent when one considers the metallicity dependence of \LcoOneZero\ shown in Figure~\ref{fig:Lco10_z0_z1_z2_z3_metallicity}. A well-established correlation exists between the stellar mass and metallicity of galaxies in both observations and simulations, known as the mass–metallicity relation (MZR) \cite{Lequeux_1979_MZR, Kinman_Davidson_1981_MZR, Garnett+2002_MZR, Tremonti_2004_MZR}. Fixing a galaxy's stellar mass is then tantamount to fixing the gas metallicity, which in turn tends to fix \LcoOneZero.

The small differences in the median \LcoOneZero\ with redshift could be attributed to the evolution of the MZR with redshift in \firebox\ \citep{bassini2024_MZR_firebox}, yet the decrease in the metallicity at higher redshifts is compensated by the increased SFRs at the higher redshifts at the fixed stellar mass \citep{Feldmann_2023}. As a result \LcoOneZero\ stays within the same order of magnitude as at $\rm z=0$.

In the remainder of this section, we compare the \fireTwo\ predictions against the observations as a function of stellar mass. At higher stellar masses our \fireTwo\ predictions do seem to match well with the observations. At $\rm z=0$, our \fireTwo\ predictions are yielding a slightly faint, although the large number of upper limits suggests this may simply reflect the xCOLD GASS sensitivity limit which deteriorates with increasing stellar mass. Below $\rm \Mstar = 10^{10}\,M_\odot$, however, our predictions become progressively fainter than the xCOLD GASS detections. This discrepancy arises because the faintest \LcoOneZero\ that xCOLD GASS can detect is comparable to the median \LcoOneZero\ of our \fireTwo\ galaxies at these masses. A more sensitive KLM survey, however, continues to detect fainter CO emitters, closer to our predictions at lower stellar masses. 

Beyond the offset in between xCOLD GASS and \fireTwo\ predictions, the two also differ in slope. The \fireTwo\ predictions exhibit an \LcoOneZero--\Mstar\ relation that is nearly twice as steep as that inferred from the xCOLD GASS detections. Over $\rm 10^{9.5} < \Mstar/M_\odot < 10^{10.5}$, the xCOLD GASS detections give a slope of $\rm d\log\LcoOneZero/d\log\Mstar \approx 0.8$, whereas \fireTwo\ yields 1.58 over the same range. We discuss possible origins of this difference in the power law relation in Section~\ref{sec:discussion_metallicity}.

\subsection{\LcoOneZero\ Luminosity Function at z $\geq 2$ \label{sec:LF_at_z23}}
\begin{figure}[ht!]
    \centering
    \includegraphics[width=0.99\linewidth]{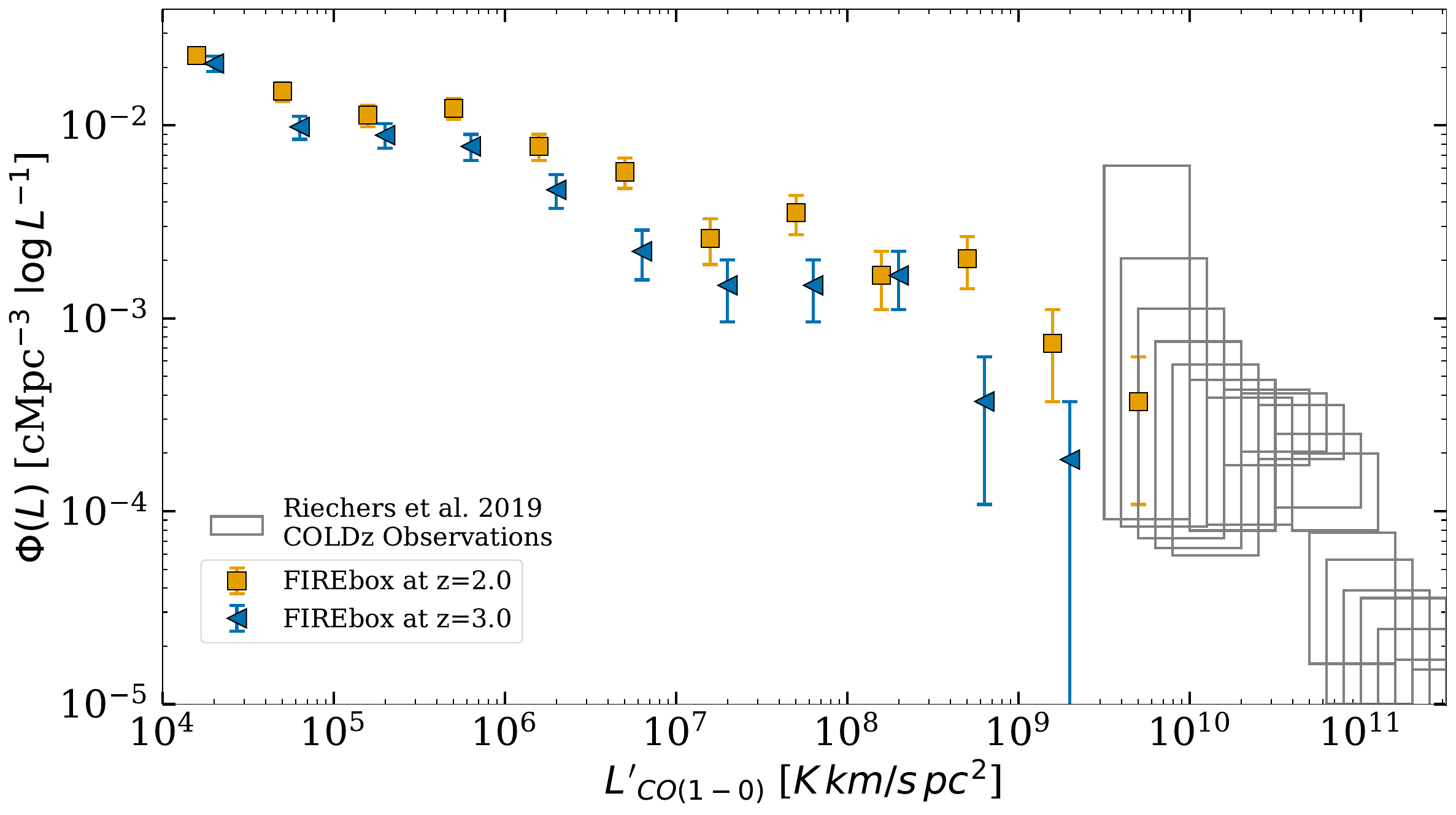}
    \caption{Comparison of the CO(1-0) luminosity function for \firebox\ galaxies at $z=2$ and $z=3$ with the COLDz observations \citep{Riechers_2019_coldz}. The COLDz measurements are shown as open green boxes, while the \firebox\ predictions are shown as filled yellow boxes and blue triangles for $z=2$ and $z=3$, respectively. The predicted \LcoOneZero\ values of the \firebox\ galaxies at $z=2$ and $z=3$ are almost all below the COLDz detection limits, with a single $z=2$ values exceeding the lower limit of the COLDz results.}
    \label{fig:co10_lf_z23_observations}
\end{figure}

We compare the CO(1-0) luminosity functions of the central halos in the \firebox\ halos at $z=2$ and $z=3$ with the COLDz survey at $2 < z < 3$ in Figure~\ref{fig:co10_lf_z23_observations}. 

At $z \geq 2$, the most luminous \firebox\ galaxies reach $\LcoOneZero \sim 6 \times 10^9~\obsunits$ at $z=2$ and only $\sim 2 \times 10^{9}~\obsunits$ at $z=3$. The COLDz survey, in contrast, is complete only down to $\LcoOneZero \approx 3 \times 10^9~\obsunits$. At $z=3$ the two therefore do not overlap at all, and at $z=2$ they overlap only over a narrow range.

This limited overlap reflects the complementary nature of the two datasets. At $z=3$, the simulated \firebox\ volume is $10{,}793~\mathrm{cMpc}^3$, whereas the COLDz survey covers a cosmic volume of $151{,}231~\mathrm{cMpc}^3$ when only the CO(1-0) line is considered. \firebox\ is designed to resolve typical galaxies with high resolution rather than to capture the rarest systems, and it therefore contains only a handful of the highly luminous CO emitters that dominate the COLDz detections. At the same time, the line sensitivity of COLDz limits its measurements to the bright end of the CO luminosity function, so the survey cannot probe the fainter and far more numerous population of CO emitters that \firebox\ simulation predicts plenty. As a result, the two luminosity functions overlap only over a narrow range in \LcoOneZero, where the survey approaches its detection limit. Taken together, \firebox\ and COLDz trace largely complementary regimes, and a more complete comparison will require both larger simulated volumes at comparable resolution and deeper CO surveys that reach lower luminosities.

As noted earlier, the \firebox\ cosmological volume is well suited to constraining the faint end of the LF, where observational surveys are incomplete. We therefore use it to probe the LF slope in the lowest luminosity bins, where our predictions show a shallow faint end. Between $10^{4}$ and $10^{9}\,\obsunits$, the number density at $\rm z=3$ falls only from $\approx2\times10^{-2}$ to $\rm \approx4\times10^{-4}\,Mpc^{-3}\log(\LcoOneZero)^{-1}$. At $\rm z=2$, the decline is even milder, reaching only $\rm \approx10^{-3}\,Mpc^{-3}\log(\LcoOneZero)^{-1}$ over the same range. These correspond to faint end slopes of $\rm d\log\phi/d\log\LcoOneZero \approx -0.25$ at $\rm z=2$ and $\approx -0.33$ at $\rm z=3$.

\section{Discussion \label{sec:discussion}}

Section~\ref{sec:results_and_comparison_with_observations} is intended to establish how far the CO luminosities predicted by our pipeline can be trusted, and over what range of galaxy properties. This matters because the predictive models in \LcoOneZero\ as a function of halo properties, which will be developed in \citetalias{Paper3}, will be applied to halo populations well beyond the reach of current observations to make mock maps for the line intensity mapping experiments. Where the simulated and observed samples overlap, namely massive star-forming galaxies at $z=0$ and the dwarfs in the KLM compilation, our predictions reproduce both the mean \LcoOneZero--SFR relation and its scatter. Outside this regime the comparison is largely untested rather than failed. This covers faint, metal-poor galaxies below survey sensitivity limits. The former is precisely where our predictions depart most from empirical models. In the remainder of this section we discuss the metallicity and stellar mass dependence that drives this departure, the systematic uncertainties in our predictions, and the implications for CO LIM.

\subsection{The Metallicity and Stellar Mass Dependence \label{sec:discussion_metallicity}}

The most consequential trend in our results is the steep dependence of CO luminosity on gas-phase metallicity. This dependence is not imposed anywhere in our pipeline. Neither the simulations nor the \cloudy\ models are calibrated to CO observations, and the trend emerges from the treatment of dust shielding and CO photodissociation in \cloudy. The trend nonetheless agrees qualitatively with the observations. The KLM compilation and the xCOLD GASS observations \citep{Saintonge_2017} discussed in Section~\ref{sec:Lco10_sfr_metallicity_at_z0} imply a steeply superlinear relation between CO luminosity and oxygen abundance, and \citet{Cicone_2017_allsmog} find a comparable steepness in the ALLSMOG sample of 88 nearby low-mass star-forming galaxies. 

There is nonetheless a systematic uncertainty that must be kept in mind. A quantitative comparison in Figure~\ref{fig:Lco10_vs_sfr10Myr_galaxyType_z0_colorBarLogOH} is limited by how metallicity is measured. Our metallicities are read directly from the gas particles, whereas observed abundances are inferred from ratios of collisionally excited lines (CLEs), and the two are not the same quantity. Furthermore, metallicities inferred from CLEs are often subject to large systematic uncertainties of the order of 0.8 dex \citep{Tremonti_2004_MZR, kewleyMetallicityCalibrationsMassMetallicity2008, Moustakas_2010_krumholzCompiledData}. This introduces yet another source of uncertainty into our comparisons and is likely the reason our predictions are fainter than the observations at the same SFR and metallicity.

In Section~\ref{sec:Lco10_dependence_on_Z_Mstar_z} we show that the initial decrease in the median \LcoOneZero\ at fixed \sfrHundredMyr\ in the \fireTwo\ predictions at increasing redshift can be partly explained by the overall decrease in metallicity at higher redshift. \fireTwo\ galaxies at different redshifts but with the same SFR and metallicity produce similar \LcoOneZero. These two quantities likely control the gas temperature and the shielding, which are the two key ingredients that determine CO emission.

In Figure~\ref{fig:Lco_vs_Mstar_redshifts} we plot \LcoOneZero\ as a function of stellar mass. At fixed stellar mass, our \fireTwo\ predictions show no statistically significant dependence on redshift. Since stellar mass correlates tightly with gas-phase metallicity, galaxies of the same stellar mass have roughly the same metallicity. Because \LcoOneZero\ correlates tightly with metallicity, it is not surprising that \Mstar\ correlates equally well with \LcoOneZero\ at a given redshift. However, the \firebox\ simulations show some evolution in the mass--metallicity relation \citep{bassini2024_MZR_firebox}, in which galaxies at higher redshifts have slightly lower metallicities. We would therefore naturally expect \LcoOneZero\ to decrease with increasing redshift at fixed stellar mass, yet we see no such decrease. Why, then, does the lower metallicity at higher redshift not produce an offset in \LcoOneZero?

It is true that at the same stellar mass, the metallicities of the \fire\ galaxies at higher redshifts are slightly lower than those of their counterparts at $z=0$. However, the same galaxies have higher SFRs at higher redshift. These two competing factors appear to offset each other, producing roughly the same \LcoOneZero\ at a given stellar mass at all redshifts. We will further investigate the correlations between galaxy properties and \LcoOneZero\ in \citetalias{Paper3}.

\subsection{Sources of systematic uncertainty \label{sec:discussion_systematics}}

Our predictions are subject to three classes of systematic uncertainty. These can be grouped as those inherited from the simulated galaxy population, those introduced by the post-processing, and those arising how we select our sample. The post-processing assumptions are described in full in \citetalias{Paper1}. Here we summarize them, note which results of this paper each could affect, and indicate the direction of the bias where it is known.

\subsubsection{The simulated galaxy population.} 
All of our galaxies are simulated with \fireTwo\ physics, so our results inherit whatever biases its star formation and feedback prescriptions carry, and we cannot assess these from within our own sample.

The \fireTwo\ predictions exhibit an \LcoOneZero--\Mstar\ relation that is nearly twice as steep as that inferred from the xCOLD GASS detections. We suggest that this difference is closely linked to the mass-metallicity relation (MZR) of the two samples, whose slope we denote $s_{\rm Z} \equiv \mathrm{d}(\twelvePlusOH)/\mathrm{d}\log\Mstar$.

The \firebox\ galaxies \citep{Feldmann_2023, bassini2024_MZR_firebox} and the \core\ \fireTwo\ galaxies at $5 < z < 12$ \citep{Marszewski2024_mzr_fire2} broadly reproduce the observed MZR, although the individual \fireTwo\ suites yield slightly different slopes. The \core\ \fireTwo\ galaxies follow $s_{\rm Z} \approx 0.37$ at $z=5$-12 \citep{Marszewski2024_mzr_fire2}, while galaxies with $\Mstar \lesssim 10^{10}\,\rm M_\odot$ in the \firebox\ cosmological volume follow a somewhat steeper relation with $s_{\rm Z} \approx 0.6$ \citep{bassini2024_MZR_firebox} and gets flatter with increasing stellar mass.

At $z=0$, the oxygen abundances in \firebox\ lie slightly above observational estimates at $\Mstar \gtrsim 10^{10}\,\rm M_\odot$ and slightly below them at $\Mstar \lesssim 10^{7.5}\,\rm M_\odot$ \citep{Feldmann_2023, bassini2024_MZR_firebox}. Taken together, these offsets mean that the simulated mass--metallicity relation (MZR) is somewhat steeper than the observed one. Using our adopted metallicity, $\twelvePlusOH_{R50}$, we find $s_{\rm Z} \approx 0.5$ for the full \fireTwo\ sample at $\Mstar < 10^{10}\,\rm M_\odot$, compared with $s_{\rm Z} \approx 0.3$ for the xCOLD GASS sample over the same mass range. Our slope is more similar to that of \firebox\ cosmological volume because our simulated galaxies used in this work are dominated by \firebox\ in number. 

This difference in MZR slope propagates directly into the \LcoOneZero--\Mstar\ relation, since
\begin{equation}
\frac{\mathrm{d}\log\LcoOneZero}{\mathrm{d}\log\Mstar} \approx \frac{\mathrm{d}\log\LcoOneZero}{\mathrm{d}(\twelvePlusOH)} \times s_{\rm Z}.
\end{equation}
Because our predicted \LcoOneZero--\twelvePlusOH\ relation has a slope similar to the observed one (Figure~\ref{fig:Lco_10_z0_z1_z2_z3_sfr100Myr}), the first factor is roughly the same in both cases. The steeper \LcoOneZero--\Mstar\ relation we predict could therefore arise from the second factor, namely the MZR, whose slope in the simulations is about 1.7 times steeper than in the xCOLD GASS data. 

We note, however, that the slope of the MZR at low stellar masses is sensitive to the adopted metallicity diagnostic, aperture, and sample selection, that the observed relations themselves carry large systematic uncertainties, and that simulated metallicities are measured directly from the gas whereas observational estimates rely on line calibrations. This offset therefore need not reflect a genuine physical discrepancy between the simulations and observations.

A steeper MZR could also imply that our metallicities at lower stellar masses could be slightly low than observed ones. Because \LcoOneZero\ decreasing very steeply with decreasing metallicity, even a slight offset in the metallicity of the simulated galaxies could result overly pessimistic \LcoOneZero\ predictions at low stellar masses.

The KLM compilation, on the other hand, gives $\mathrm{d}\log\LcoOneZero/\mathrm{d}\log\Mstar \approx 1.4$, which lies between the xCOLD GASS value and our \fireTwo\ prediction. Since KLM compilation extend to fainter CO luminosities than xCOLD GASS, the steeper slope they yield suggest that the shallower xCOLD GASS slope partly be a a consquence of limited sensitivity. Deeper CO observations of low-mass galaxies will be needed to test this.

\subsubsection{Cloud geometry} 
We model each gas particle as a one-dimensional, constant-density, plane-parallel slab illuminated on one face, with a stopping length equal to its smoothing length. The emitting gas, in contrast, is three-dimensional, clumpy, and has no well-defined boundaries \citep{Dame2001_COdata_for_MivilleDeschenes}. 

We prefer this geometry to the spherical one for two reasons. First, only a small fraction of Milky Way molecular clouds are forming stars at any given time \citep{Miville-Deschenes_2017_MW_paper, Chevance2020_tuningFork_phangs}, so most of the CO emission arises in gas heated from outside by the ambient interstellar radiation field rather than by embedded sources. Because the young stars that supply the dissociating far-ultraviolet photons have a vertical scale height comparable to that of the molecular gas, treating this field as incident on one face of the cloud is a reasonable approximation. 

We tested our approximations against observed Milky Way (MW) molecular clouds in \citetalias{Paper1}, finding a good agreement MW data and our \cloudy\ predictions for the clouds with $\rm A_v\geq1$. That our galaxy integrated luminosities agree with observations at $z=0$ could suggest that errors at the cloud scale partly cancel when summed over a galaxy, but this still does not show that this cancellation holds in general for all \fireTwo\ galaxies.

\subsubsection{Radiation field and cosmic rays} 

Although \skirt\ returns the full local radiation field, we pass only its FUV normalization $\rm G_{\rm FUV}$ to \cloudy\ and adopt the MW spectral shape \citep{Black_1987_isrf} for \cloudy\ simulation that we interpolated for our gas particles. Therefore we assumed MW spectral shape for all of our gas particles whereas in reality the shape of the SED can vary from one location to another in the galaxy. 

Furthermore, because $\rm G_{\rm FUV}$ is integrated over the full 6-13.6 eV band, it does not constrain how the energy is distributed within that band. Younger and more metal-poor stellar populations produce harder spectra, and metal-poor galaxies contain less dust to redden the field. This is partly accounted for in the $\rm G_{\rm FUV}$; however, the adopted constant does not separately follow the 11.1-13.6 eV photons that photodissociate CO. Therefore, at high redshifts, the 11.1 - 13.6 eV photons can make up a larger fraction of the band than the assumed MW spectral shape at the same $\rm G_{\rm FUV}$. This would lower the CO abundance in low mass and high redshift galaxies, so the redshift decline in Figure~\ref{fig:Lco_10_z0_z1_z2_z3_sfr100Myr} may be underestimated.

Similarly, we adopt the Milky Way cosmic-ray ionization rate in all of our \cloudy\ simulations. Galaxies with higher star formation are expected to have substantially higher cosmic ray production efficiently destroying CO molecules deep inside the well-shielded gas \citep{bisbas_2015_co_dark_clouds}. Our predicted \LcoOneZero\ for the most actively star forming galaxies may therefore be too high. This is also seen in \citet{Keating_2020}, who found for MW analogues in \fireTwo\ that a tenfold increase in the ionization rate noticeably lowers CO abundance and therefore \LcoOneZero. Cosmic rays also heat the gas, which raises the excitation temperature and may partly offset this reduction. 

\subsection{Dust and metal scaling}

We assume a constant dust to metal ratio of 0.4 and a single grain mixture at all metallicities and redshifts. Observations suggest that the dust to gas ratio falls more steeply than linearly with metallicity below $\sim0.1$-$0.2\,Z_\odot$ \citep{RemyRuyer2014_dustToGas}. If so, we overestimate the dust shielding in our most metal-poor galaxies and overpredict their CO emission. 

Most consequentially, in our \cloudy\ runs we use predetermined metal abundances and scale all of them as a function of the total metal mass fraction of each gas cell, normalized to solar units using the fiducial value of $\rm Z_\odot = 0.0134$. In other words, we do not use the individual oxygen and carbon abundances of our gas cells. Given that our CO luminosities are sensitive to metallicity, this simplification is likely one of the dominant systematics in our predictions, and one that we have not tested. Doing so would require extending the \cloudy\ lookup table to a larger number of dimensions, making training and interpolation more computationally demanding. We leave this for future work. 

\subsection{Line radiative transfer}

\cloudy\ accounts for the optical depth of the CO lines within each simulated slab, but we sum the luminosities of individual gas particles without doing another line radiative transfer between gas elements. We therefore neglect absorption of CO emission by foreground gas at the same velocity, as well as any dependence of the emergent luminosity on viewing angle. In \citetalias{Paper1} we argued that this is a small effect for MW-like galaxies. For CO photons from one cloud to be absorbed by another, the two clouds must overlap both on the sky and in velocity. 

In the Galactic plane, the number of clouds along a random line of sight peaks at one, and even the rarer sight lines that cross six or more clouds show them well separated in velocity \citep{Miville-Deschenes_2017_MW_paper}. Self-extinction therefore appears to have little impact on our predicted luminosities. This argument is weaker for the compact, gas-rich, high-surface-density galaxies that become common at $z\geq2$. There, a line of sight is more likely to cross several clouds that also overlap in velocity, so our summed luminosities may yield overestimates. 

The effect is compounded by our definition of the turbulent velocity, which includes rotation, shear and bulk motions as well as true microturbulence. The resulting line widths are broader than the local ones, which lowers the CO optical depth in each slab and raises the luminosity that escapes. A full three-dimensional line radiative transfer calculation for a representative subsample of our galaxies, for example with \texttt{RADMC-3D} \citep{Dullemond_2012_RADMC3D}, would test both effects directly, though in that case an accurate determination of the CO gas temperature and abundances is required to be pre-known or computed via another code.

\subsection{Sample selection} 
Because we work with logarithmic quantities, we exclude the $\approx5\%$ of galaxies with no star formation in the past $100\Myr$ or no gas within the half-light radius. These are systematically the faintest CO emitters, so excluding them likely leads us to underestimate the scatter in \LcoOneZero\ at fixed stellar mass. 

At the other extreme, our sample contains few massive halos at high redshift. Even the virial mass of the most massive \massivefire\ halos reach only $\approx3\times10^{12}\,M_\odot$ at $z=2$ and $\approx1.6\times10^{12}\,M_\odot$ at $z=3$, below the most massive halos expected at these redshifts. Our predictions therefore does not probe the very massive
end of the halo mass function well.

\subsection{Implications for CO line intensity mapping}
\label{sec:discussion_implications_co_LIM}

Subject to the caveats mentioned above, our results stands out as consequential for LIM forecasting. At fixed SFR, the CO luminosities of our \fireTwo\ galaxies falls by roughly an order of magnitude per unit redshift. We have shown that this decline is more closely related to the reducing metallicity of the galaxy populations at higher redshifts. In Figure~\ref{fig:Lco10_z0_z1_z2_z3_metallicity} we show galaxies at higher redshifts with similar metallicities and SFRs produce comparable CO luminosities.

Most response functions (or distribution functions) currently used in CO LIM forecasts connect halo mass to CO luminosity through an intermediate star formation tracer \citep{Lidz_2011_coModelling, pullen_2013_coModelling, Li_2016_coModelling, Keating_2020_coModelling, dongwoo_2022_comap_earlyscience5}. Such a chain should inherit the redshift behaviour of the tracer, yet these models are not adequately informed by the chemical evolution of the galaxies with exceptions such as LIMFAST \citep{masRibas2023_LIMfast1, jasonSun2023_LIMfast2_coModelling}. Our predictions indicate that, at fixed star formation rate, CO luminosity decreases toward higher redshift. This is because galaxies at earlier times are less chemically enriched and have lower stellar masses than their lower redshift counterparts. The fact that most of the response functions does not include metallicity information in their models, they likely contain large systematic uncertainties. This offers a physical reasoning of at least part of the several orders of magnitude spread among published CO models especially at the lowest halo masses \citep{Breysse_2016_highredshift_COmocks, dongwoo_2024_COMAP_season2}, which current Carbon Monoxide Mapping Array Project (COMAP, \cite{Cleary_2022_comap_earlyScience1}) have only begun to narrow. We leave a model-by-model comparison with the literature to \citetalias{Paper3}, which develops the \LcoOneZero-halo mass relation used for that comparison.

The broader point we would like emphasize is the complementary nature of this work to empirically calibrated models. The currently used models in the literature are anchored in bright, massive and relatively metal-rich galaxies as they are the ones current surveys can detect within their sensitivity limit. The calculations presented here samples a population largely inaccessible to those surveys such as the low-mass, metal-poor galaxies that dominate our sample and our Universe by number. 

Taken together, these approaches constrain \LcoOneZero\ across complementary regimes, with empirical calibrations anchored in the massive, metal-rich population and our simulation-based framework extending predictions into the low-mass, metal-poor regime that remains largely unconstrained by current observations.

\section{Conclusion \label{sec:conclusion}}

We have combined \fireTwo\ simulations, \skirt\ radiative transfer, and \cloudy\ line emission calculations to predict \LcoOneZero\ for 765, 850, 887 and 915 simulated \fireTwo\ galaxies at $z=0$, 1, 2 and 3, spanning 5.6 orders of magnitude in stellar mass and 3.1 orders of magnitude in gas-phase metallicity. This paper, second in series, compares those predictions against CO observations from $z=0$ to $z=3$. The pipeline and it's validation against MW clouds, molecular gas masses and CO abundances are presented in \citetalias{Paper1} and the response functions, which can be used to rapidly paint the dark matter halo catalogs, built on the results reported here are presented in \citetalias{Paper3}.

In this paper we compared our \LcoOneZero\ and \LcoOneZero\ predictions against both local and high redshift observations. At $z=0$ in the regime where simulations and surveys overlap the agreement is good. Our \LcoOneZero\ predictions match the xCOLD GASS detections \citep{Saintonge_2017} and KLM compilation \citep{Krumholz_2011_lowZ_z0_observations} across the full range of SFRs with comparable scatter as shown in Figure~\ref{fig:Lco21_vs_sfr10Myr_galaxyType_z0}. Our \fireTwo\ predictions are recreate a similar \LcoOneZero-metallicity slope observed in KLM compilation.

The predicted CO(1-0) luminosity function from the full \firebox\ volume is predicting slight excess CO emitters compared to xCOLD GASS data above $\LcoOneZero \approx 2\times10^{8}\,\obsunits$, but the tension reaches to $\approx4\sigma$ near the sensitivity limit of the xCOLD GASS survey for highest stellar masses.

At higher redshifts our \fireTwo\ predictions are showing much steeper correlation with $\rm SFR$, while the observed relation is fairly flat. Our \fireTwo\ predictions exhibit substantial scatter and generally lie below the observations at fixed SFR. However, the detection limits make it difficult to determine whether this discrepancy arises from observational incompleteness due to undetected galaxies or from systematically underpredicted line luminosities by our post-processing framework. Albeit, it is true that the observed and simulated galaxies differ in their star formation efficieny and the depletion time, therefore in Section~\ref{sec:Lco21_z1_fgas_tdep_Mstar_sfr_matched} we compare the galaxies that occupy the same range of stellar mass, SFR, gas fractions and depletion times and found that the disagreement between \fireTwo\ predictions and observations are driven by the $\alpha_{CO}$ factor which is itself strongly depends on the metallicity. Our \fireTwo\ predictions with $\alpha_{CO}\approx5\,M_\odot(\obsunits)^{-1}$ agree well with the data, while the ones that scatter to much lower \LcoTwoOne\ has more than order of magnitude larger $\alpha_{CO}$ driven by their low metallicities. Therefore outside the region where observations are not sensitive the comparison is largely untested rather than failed. Our galaxies extend to \LcoOneZero\ and \LcoTwoOne\ well below current detection limits, and the disagreements especially at low SFR or low metallicity may reflect a faint population that the surveys cannot robustly detect.

Furthermore, we find no evidence that the results depend on numerical resolution, e.g. the \core\ \fireTwo\ and \tripleLatte\ predictions are consistent with the lower-resolution \firebox\ galaxies at all SFRs.

Our most consequential result is the decline of \LcoOneZero\ with redshift at fixed SFR, amounting to roughly an order of magnitude per unit increase in redshift. We have shown in Figure~\ref{fig:Lco10_z0_z1_z2_z3_metallicity} that this decline is driven by metallicity, e.g. the galaxies at different redshifts but with comparable SFR and metallicity produce comparable CO luminosities. In our framework, this dependence emerges directly from the \cloudy\ calculations, and in \citetalias{Paper1} we show that it most likely arises from the reduced dust shielding of CO molecules at low metallicity environments. This interpretation is consistent with the well increase of $X_{\rm CO}$ at low metallicity environments \citep{Israel_1997, Bolatto_2013, Feldmann2012_Xco_paperI, Madden_2020}. Our \fireTwo\ predictions yield a tight correlation between stellar mass and \LcoOneZero, which likely reflects the tight MZR. More interestingly, at fixed stellar mass we find no statistically significant redshift dependence, even though the MZR evolves with redshift. We attribute this to the higher SFRs of galaxies at higher redshift at the same stellar mass, which compensate for the decrease in \LcoOneZero\ expected from their lower metallicities.

If the strong metallicity dependency we observe in our \fireTwo\ calculations is real, the fear is that the cosmic LIM signal at high-$z$ could be much lower than its counterpart at $\rm z=0$ at the same halo mass due to metallicity dependent suppression. However, at the same time this can be compensated with the increased SFRs at higher redshifts for a given halo mass. We provide our best model estimates in \citetalias{Paper3} for the cosmic LIM signal at high-$z$, and we defer further discussion our \citetalias{Paper3}. 

Testing our predictions will require simulations with different feedback prescriptions, larger volumes to sample the cosmic scatter and rare galaxies, higher resolution to resolve molecular cloud structure, and a larger number of zoom-in simulations of massive halos. Above all, however, it will require more sensitive CO observations of dwarf and metal-poor galaxies, which would provide the most direct test in the regime where empirical calibrations are least constrained.

\vspace{5mm}

\begin{acknowledgments}
This research was supported in Canada by NSERC and CIFAR. AW received support via NSF CAREER award AST-2045928. The \skirt\ and \cloudy\ calculations used to post-process the \fireTwo\ simulations were carried out on the Niagara supercomputer and its successor, Trillium, at the SciNet HPC Consortium, on the Fir supercomputer at the Cedar Supercomputing Centre at Simon Fraser University, operated through the Digital Research Alliance of Canada, and on the Sunnyvale computing cluster at CITA. We parallelized the runs across many nodes using GNU Parallel \citep{GNUparallel} while running on Trillium. SciNet receives funding from the Canada Foundation for Innovation under the auspices of the Digital Research Alliance of Canada, the Government of Ontario, the Ontario Research Fund--Research Excellence, and the University of Toronto. 

The first author (D.T.) is grateful to John Dubinski, MJ Huang, Jay Li and Daniel Xiao for their tireless work maintaining these resources. DT thank Nathan J. Carlson for helpful discussion and Guochao Sun for the comments on the paper. D.T. also thanks the Digital Research Alliance of Canada team and the staff of the Cedar Supercomputing Centre at Simon Fraser University for their support and for maintaining the Fir and Trillium supercomputers. This work used snapshots of \fireTwo\ simulations run on XSEDE, supported by NSF grant ACI-1548562; Blue Waters, supported by the NSF; Frontera allocations AST21010 and AST20016, supported by the NSF and the Texas Advanced Computing Center; and Pleiades, through the NASA HEC program via the NAS Division at Ames Research Center.

In preparing this paper, D.T., made use of three large language model (LLM) based assistants: Claude (Anthropic), ChatGPT (OpenAI), and Gemini (Google). They were employed to edit the writing for grammar and readability, to support the development and debugging of the Python code behind our analysis and figures. 

The authors carefully checked, tested, and validated all text, code, and analysis that were produced with the assistance of these tools. No AI tools were used to generate the simulation data. The authors alone are responsible for the scientific results, their interpretation, and the conclusions presented in this work.
\end{acknowledgments}

\begin{contribution}
DT led the project, performed all of the coding, analysis, and wrote the manuscript. NM and RB supervised DT. RB, NM and DC contributed substantially to the analysis of the results. All authors provided comments on the manuscript.
\end{contribution}

\software{astropy \citep{2013A&A...558A..33A,2018AJ....156..123A},
SKIRT \citep{Baes_2011, Baes_2015, Camps_2015, Camps_and_baes_2020_skirt9},
          Cloudy \citep{2013RMxAA..49..137F, Ferland_2017, cloudy_version_c25_release},
          SciPy \citep{scipy_2020},
          Meshoid (\url{https://github.com/mikegrudic/meshoid}),
          GNU Parallel \citep{GNUparallel},
          GIZMO \citep{Hopkins_2015},
          ROCKSTAR \citep{Behroozi_2013},
          AHF \citep{Gill_2004, Knollmann_2009},
          MUSIC \citep{Hahn_2011},
          STARBURST99 \citep{Leitherer_1999}
  }

\appendix

\section{KLM galaxy sample}
\label{sec:krumholzCompiledData}

We list the stellar masses for the galaxies in the KLM sample  \citep{Krumholz_2011_lowZ_z0_observations} in Table~\ref{table:CO_SFR_metallicity_krumholz2011}.

\begingroup
\setlength{\LTcapwidth}{\textwidth}
\begin{longtable*}{lcccccccc}
\caption{CO luminosities, star formation rates, metallicities, and stellar masses of nearby galaxies compiled from various literature sources. The \LcoOneZero, $SFR$, and metallicity values are taken directly from \cite{Krumholz_2011_lowZ_z0_observations}. Stellar masses are taken from the references indicated in the final column. The abbreviations used are:
B20 = \cite{Belland_2020_krumholzCompiledData_stellarMass};
Br19 = \cite{Bresolin_2019_krumholzCompiledData_stellarMasses};
BV16 = \cite{Baes_2016_krumholzCompiledData_stellarMass};
C10 = \cite{Calzetti_2010_sfr_calibration};
Co03 = \cite{corbelli_2003_krumholzCompiledData_stellarMass};
D84 = \cite{Dufour_1984_krumholzCompiledData};
E08 = \cite{Engelbracht_2008_krumholzCompiledData};
F08 = \cite{Fukui_2008_krumholzCompiledData};
H03 = \cite{Helfer_2003_krumholzCompiledData};
H04 = \cite{Heyer_2004_krumholzCompiledData};
H09 = \cite{Harris_2009_krumholzCompiledData};
I97 = \cite{Israel_1997};
J03 = \cite{jarrett_2003_ic10_Mstar};
J12 = \cite{Johnson_2012_krumholzCompiledData_stellarMass};
K03 = \cite{Kennicutt_2003_krumholzCompiledData};
K13 = \cite{Karczewski_2013_krumholzCompiledData_stellarMass};
K16 = \cite{kijeong_2016_krumholzCompiledData_stellarMass};
Ka15 = \cite{Kang_2015_krumholzCompiledData_stellarMass};
L03 = \cite{Lee_2003_krumholzCompiledData};
L06 = \cite{Leroy_2006_krumholzCompiledData};
L09 = \cite{Leroy_2009_krumholzCompiledData};
L79 = \cite{Lequeux_1979_MZR};
LS11 = \cite{Lopez_Sanchez_2011_krumholzCompiledData_stellarMass};
Lu04 = \cite{Lundgren_2004_krumholzCompiledData_stellarMasses};
M06 = \cite{Mizuno_2006_krumholzCompiledData};
M97 = \cite{Martin_1997_krumholzCompiledData};
MA10 = \cite{Marble_2010_krumholzCompiledData};
MO10 = \cite{Moustakas_2010_krumholzCompiledData};
R08 = \cite{Rosolowsky_2008_krumholzCompiledData};
S11 = \cite{Skibba_2011_krumholzCompiledData_stellarMasses};
Sa22 = \cite{Salvestrini_2022_krumholzCompiledData_stellarMasses};
T98 = \cite{Taylor_1998_krumholzCompiledData};
vdM09 = \cite{vdm_2009_krumholzCompiledData_stellarMass};
W04 = \cite{Wilke_2004_krumholzCompiledData};
Y95 = \cite{Young_1995_krumholzCompiledData};
Z13 = \cite{Zhao_2013_krumholzCompiledData_stellarMass}. $^{a}$~S11 also list M33 (as NGC\,0598), with a lower stellar mass of $\log M_{star} = 8.86 \pm 0.10$; we adopt the Co03 value.}
\label{table:CO_SFR_metallicity_krumholz2011}\\
\hline\hline
Galaxy & $\log L'_{\mathrm{CO(1-0)}}$ & CO Ref. & $\log SFR$ & SFR Ref. & 12 + $\log(\mathrm{O/H})$ & Metallicity Ref. &  $\log {M}_{star}$ & $M_{star}$ Ref. \\
 & (K km s$^{-1}$ pc$^2$) & & (M$_\odot$ yr$^{-1}$) & & & & (M$_\odot$) & \\
\hline
\endfirsthead

\caption[]{(continued)}\\
\hline\hline
Galaxy & $\log L'_{\mathrm{CO(1-0)}}$ & CO Ref. & $\log SFR$ & SFR Ref. & 12 + $\log(\mathrm{O/H})$ & Metallicity Ref. &  $\log {M}_{star}$ & $M_{star}$ Ref. \\
 & (K km s$^{-1}$ pc$^2$) & & (M$_\odot$ yr$^{-1}$) & & & & (M$_\odot$) & \\
\hline
\endhead

\hline
\multicolumn{9}{r}{{Continued on next page}} \\
\endfoot

\hline
\endlastfoot
SMC & $5.2 \pm 0.2$ & M06 & $-1.3 \pm 0.2$ & W04 & $8.0 \pm 0.2$ & D84; MA10 & 8.49 & vdM09 \\
LMC & $6.5 \pm 0.1$ & F08 & $-0.7 \pm 0.2$ & H09 & $8.3 \pm 0.3$ & D84; MA10 & 9.43 & vdM09 \\
IC\,10 & $6.3 \pm 0.2$ & L06 & $-1.0 \pm 0.5$ & L06 & $8.2 \pm 0.2$ & L79; L03 & 8.60 & L06; J03 \\
M33 & $7.6 \pm 0.1$ & H04 & $0.0 \pm 0.5$ & H04 & $8.3 \pm 0.2$ & R08 & $9.65 \pm 0.14$ & Co03$^{a}$ \\
II\,Zw\,40 & $6.2 \pm 0.3$ & T98 & $-0.2 \pm 0.3$ & C10 & $8.1 \pm 0.3$ & E08; C10 & 8.08 & Z13 \\
NGC\,1569 & $5.1 \pm 0.3$ & T98 & $-0.6 \pm 0.3$ & C10 & $8.1 \pm 0.2$ & M97 & 8.45 & J12 \\
NGC\,2537 & $5.5 \pm 0.3$ & T98 & $-1.1 \pm 0.3$ & C10 & $8.4 \pm 0.3$ & MA10 & & \\
NGC\,4449 & $6.8 \pm 0.3$ & Y95 & $-0.5 \pm 0.3$ & C10 & $8.3 \pm 0.2$ & M97 & 9.00 & K13 \\
NGC\,5253 & $5.8 \pm 0.3$ & T98 & $-0.3 \pm 0.3$ & C10 & $8.2 \pm 0.3$ & MA10 & $9.06 \pm 0.02$ & LS11 \\
NGC\,6822 & $5.1 \pm 0.2$ & I97 & $-2.0 \pm 0.3$ & C10 & $8.4 \pm 0.3$ & MO10 & 8.00 & B20 \\
NGC\,0628 & $8.3 \pm 0.1$ & L09 & $-0.2 \pm 0.2$ & C10 & $8.8 \pm 0.3$ & MO10 & $9.57 \pm 0.13$ & S11 \\
NGC\,0925 & $7.6 \pm 0.1$ & L09 & $-0.3 \pm 0.2$ & C10 & $8.6 \pm 0.3$ & MO10 & $9.48 \pm 0.14$ & S11 \\
NGC\,1482 & $8.8 \pm 0.3$ & Y95 & $0.5 \pm 0.2$ & C10 & $8.6 \pm 0.4$ & MO10 & $9.99 \pm 0.11$ & S11 \\
NGC\,2146 & $9.3 \pm 0.3$ & Y95 & $0.9 \pm 0.2$ & C10 & $8.7 \pm 0.3$ & E08; C10 & $10.30 \pm 0.13$ & S11 \\
NGC\,2403 & $7.1 \pm 0.3$ & Y95 & $-0.4 \pm 0.2$ & C10 & $8.6 \pm 0.2$ & MO10 & & \\
NGC\,2841 & $8.4 \pm 0.1$ & L09 & $0.1 \pm 0.2$ & C10 & $9.0 \pm 0.3$ & MO10 & $10.17 \pm 0.14$ & S11 \\
NGC\,2782 & $9.0 \pm 0.3$ & Y95 & $0.7 \pm 0.2$ & C10 & $8.6 \pm 0.4$ & E08; C10 & 10.84 & K16 \\
NGC\,2798 & $8.8 \pm 0.3$ & Y95 & $0.5 \pm 0.2$ & C10 & $8.7 \pm 0.3$ & MO10 & $10.04 \pm 0.13$ & S11 \\
NGC\,2976 & $7.0 \pm 0.1$ & L09 & $-1.0 \pm 0.2$ & C10 & $8.7 \pm 0.3$ & MO10 & $8.97 \pm 0.11$ & S11 \\
NGC\,2903 & $8.8 \pm 0.1$ & H03 & $0.3 \pm 0.2$ & C10 & $8.9 \pm 0.3$ & MA10 & 10.63 & Br19 \\
NGC\,3034 & $8.8 \pm 0.3$ & Y95 & $0.9 \pm 0.2$ & C10 & $8.8 \pm 0.3$ & MO10 & & \\
NGC\,3077 & $6.0 \pm 0.3$ & T98 & $-1.0 \pm 0.2$ & C10 & $8.6 \pm 0.3$ & MA10 & $9.29 \pm 0.07$ & S11 \\
NGC\,3079 & $9.4 \pm 0.3$ & Y95 & $0.5 \pm 0.2$ & C10 & $8.6 \pm 0.4$ & E08; C10 & 9.76 & Sa22 \\
NGC\,3184 & $8.4 \pm 0.1$ & L09 & $-0.5 \pm 0.2$ & C10 & $8.9 \pm 0.3$ & MO10 & $9.24 \pm 0.17$ & S11 \\
NGC\,3198 & $7.9 \pm 0.1$ & L09 & $-0.0 \pm 0.2$ & C10 & $8.8 \pm 0.3$ & MO10 & $9.85 \pm 0.11$ & S11 \\
NGC\,3310 & $8.2 \pm 0.3$ & Y95 & $0.9 \pm 0.2$ & C10 & $8.2 \pm 0.4$ & E08; C10 & & \\
NGC\,3351 & $8.2 \pm 0.1$ & L09 & $-0.2 \pm 0.2$ & C10 & $8.9 \pm 0.3$ & MO10 & $10.28 \pm 0.12$ & S11 \\
NGC\,3368 & $8.3 \pm 0.3$ & Y95 & $-0.4 \pm 0.2$ & C10 & $9.0 \pm 0.3$ & MA10 & & \\
NGC\,3521 & $8.8 \pm 0.1$ & L09 & $0.1 \pm 0.2$ & C10 & $8.8 \pm 0.3$ & MO10 & $10.78 \pm 0.12$ & S11 \\
NGC\,3628 & $9.2 \pm 0.3$ & Y95 & $0.3 \pm 0.2$ & C10 & $9.0 \pm 0.3$ & MA10 & 10.75 & BV16 \\
NGC\,3627 & $9.0 \pm 0.1$ & H03 & $0.2 \pm 0.2$ & C10 & $8.7 \pm 0.3$ & MO10 & $10.57 \pm 0.13$ & S11 \\
NGC\,3938 & $8.5 \pm 0.1$ & H03 & $-0.1 \pm 0.2$ & C10 & $8.7 \pm 0.3$ & E08; C10 & $9.12 \pm 0.04$ & S11 \\
NGC\,4194 & $8.9 \pm 0.3$ & Y95 & $1.1 \pm 0.2$ & C10 & $8.7 \pm 0.4$ & E08; C10 & & \\
NGC\,4214 & $6.1 \pm 0.1$ & L09 & $-1.0 \pm 0.2$ & C10 & $8.2 \pm 0.2$ & T98 & & \\
NGC\,4254 & $9.9 \pm 0.3$ & Y95 & $1.3 \pm 0.2$ & C10 & $8.8 \pm 0.3$ & MO10 & $9.61 \pm 0.15$ & S11 \\
NGC\,4321 & $9.2 \pm 0.1$ & H03 & $0.4 \pm 0.2$ & C10 & $8.8 \pm 0.3$ & MO10 & $10.36 \pm 0.11$ & S11 \\
NGC\,4450 & $8.9 \pm 0.3$ & Y95 & $-0.2 \pm 0.2$ & C10 & $8.9 \pm 0.4$ & C10; MA10 & & \\
NGC\,4536 & $8.6 \pm 0.3$ & Y95 & $0.3 \pm 0.2$ & C10 & $8.6 \pm 0.4$ & MO10 & $9.49 \pm 0.11$ & S11 \\
NGC\,4569 & $8.8 \pm 0.1$ & H03 & $-0.1 \pm 0.2$ & C10 & $8.9 \pm 0.4$ & E08; C10 & $10.38 \pm 0.12$ & S11 \\
NGC\,4579 & $9.0 \pm 0.3$ & Y95 & $0.2 \pm 0.2$ & C10 & $9.0 \pm 0.4$ & C10; MA10 & $9.96 \pm 0.23$ & S11 \\
NGC\,4631 & $8.5 \pm 0.3$ & Y95 & $0.4 \pm 0.2$ & C10 & $8.4 \pm 0.3$ & MO10 & $9.76 \pm 0.14$ & S11 \\
NGC\,4725 & $9.1 \pm 0.3$ & Y95 & $0.0 \pm 0.2$ & C10 & $8.7 \pm 0.4$ & MO10 & $10.58 \pm 0.12$ & S11 \\
NGC\,4736 & $7.9 \pm 0.1$ & L09 & $-0.4 \pm 0.2$ & C10 & $8.7 \pm 0.4$ & MO10 & $10.34 \pm 0.13$ & S11 \\
NGC\,4826 & $8.1 \pm 0.1$ & H03 & $-0.5 \pm 0.2$ & C10 & $8.9 \pm 0.3$ & MO10 & $9.99 \pm 0.12$ & S11 \\
NGC\,5033 & $9.3 \pm 0.1$ & H03 & $0.1 \pm 0.3$ & K03 & $8.7 \pm 0.3$ & MO10 & 10.57 & Br19 \\
NGC\,5055 & $8.9 \pm 0.1$ & L09 & $0.1 \pm 0.2$ & C10 & $8.9 \pm 0.4$ & MO10 & $10.76 \pm 0.12$ & S11 \\
NGC\,5194 & $9.2 \pm 0.1$ & H03 & $0.4 \pm 0.2$ & C10 & $9.0 \pm 0.3$ & MO10 & 10.56 & Ka15 \\
NGC\,5236 & $8.9 \pm 0.1$ & Y95 & $0.4 \pm 0.2$ & C10 & $9.0 \pm 0.3$ & MO10 & 10.57 & Lu04 \\
NGC\,5713 & $9.1 \pm 0.3$ & Y95 & $0.6 \pm 0.2$ & C10 & $8.7 \pm 0.4$ & MO10 & $10.07 \pm 0.11$ & S11 \\
NGC\,5866 & $8.1 \pm 0.3$ & Y95 & $-0.6 \pm 0.2$ & C10 & $8.7 \pm 0.4$ & C10; MA10 & $10.02 \pm 0.09$ & S11 \\
NGC\,5953 & $9.0 \pm 0.3$ & Y95 & $0.4 \pm 0.2$ & C10 & $8.7 \pm 0.4$ & E08; C10 & 9.99 & Sa22 \\
NGC\,6946 & $8.8 \pm 0.1$ & L09 & $0.5 \pm 0.2$ & C10 & $8.8 \pm 0.3$ & MO10 & $9.96 \pm 0.40$ & S11 \\
NGC\,7331 & $9.0 \pm 0.1$ & L09 & $0.4 \pm 0.2$ & C10 & $8.8 \pm 0.3$ & MO10 & $10.58 \pm 0.12$ & S11 \\

\end{longtable*}
\endgroup

\DeclareRobustCommand{\redshiftOne}{$\rm z\sim1$}
\section{\NoCaseChange{\LcoTwoOne} versus SFR at \redshiftOne\ for Galaxies with Matched Gas Fractions \label{sec:Lco21_z1_fgas_tdep_Mstar_sfr_matched}}

\begin{figure*}[!t]
    \centering
    \includegraphics[width=0.99\linewidth]{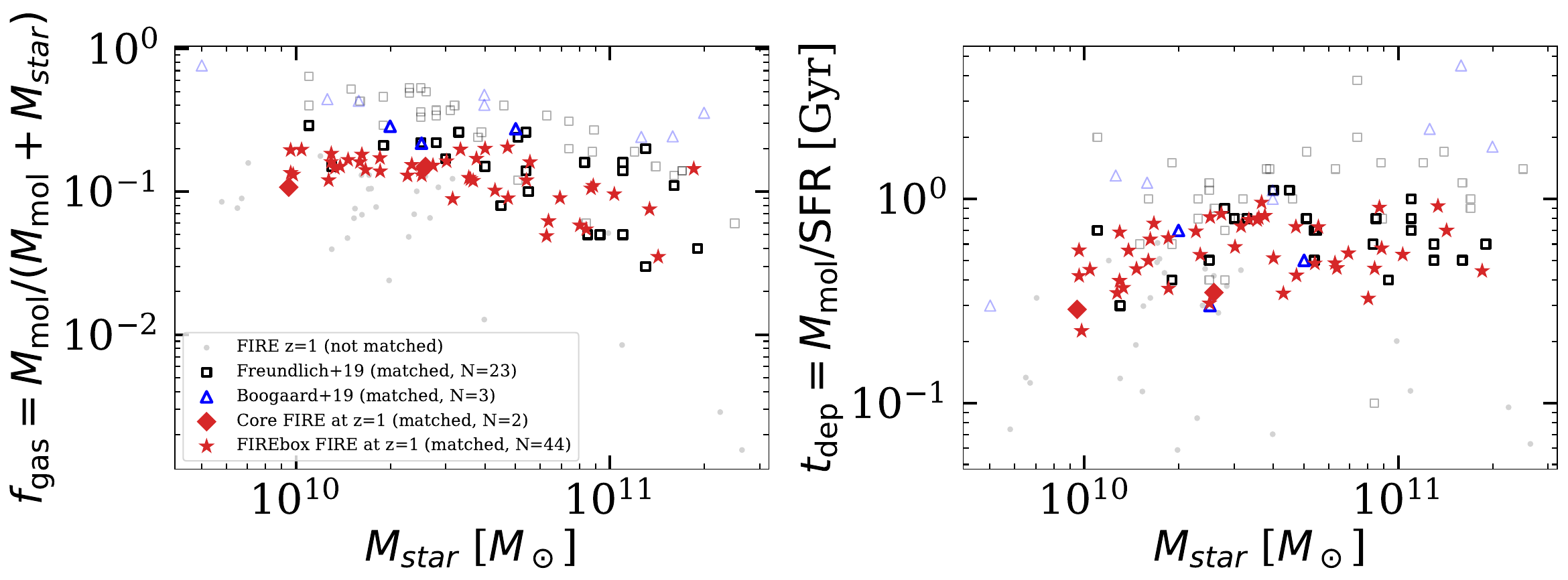}
    \caption{Molecular gas fraction, $f_{\rm gas} = M_{\rm mol}/(M_{\rm mol} + \Mstar)$ (left), and molecular gas depletion time, $t_{\rm dep} = M_{\rm mol}/{\rm SFR}$ (right), as a function of stellar mass for \fire\ galaxies at $z = 1$ and for the PHIBSS2 \citep{Freundlich_2019} and ASPECS \citep{boogaard_2019_alma_aspecs} CO(2--1) samples. Each \fire\ galaxy is paired with every observed galaxy whose $\Mstar$, $f_{\rm gas}$ and $t_{\rm dep}$ all lie within 0.2\,dex of its own. Since $M_{\rm mol} = \Mstar f_{\rm gas}/(1 - f_{\rm gas})$ and ${\rm SFR} = M_{\rm mol}/t_{\rm dep}$, the matched pairs therefore also have similar molecular gas masses and SFRs. Red stars and diamonds show the matched FIREbox (44) and Core \fire\ (2) galaxies; black squares and blue triangles show the matched PHIBSS2 (23) and ASPECS (3) galaxies. Grey points and faint symbols show the unmatched \fire\ and observed galaxies. For \fire\, $M_{\rm mol}$ is the H$_2$ mass predicted by our Cloudy models multiplied by 1.37 to include helium, and the SFR is averaged over the last 100\,Myr. The observed $M_{\rm mol}$ are those derived by the surveys from $L'_{\rm CO(2-1)}$. Before matching, the \fire\ galaxies are restricted to the $\Mstar$ and SFR ranges covered by the observations. Only the gas-poor tail of the observed samples has \fire\ counterparts.}
    \label{fig:matched_fgas_tdep_vs_Mstar_z1_sfr100Myr_Mstar0.2_fgas0.2_tdep0.2}
\end{figure*}

\begin{figure*}[!t]
    \centering
    \includegraphics[width=0.99\linewidth]{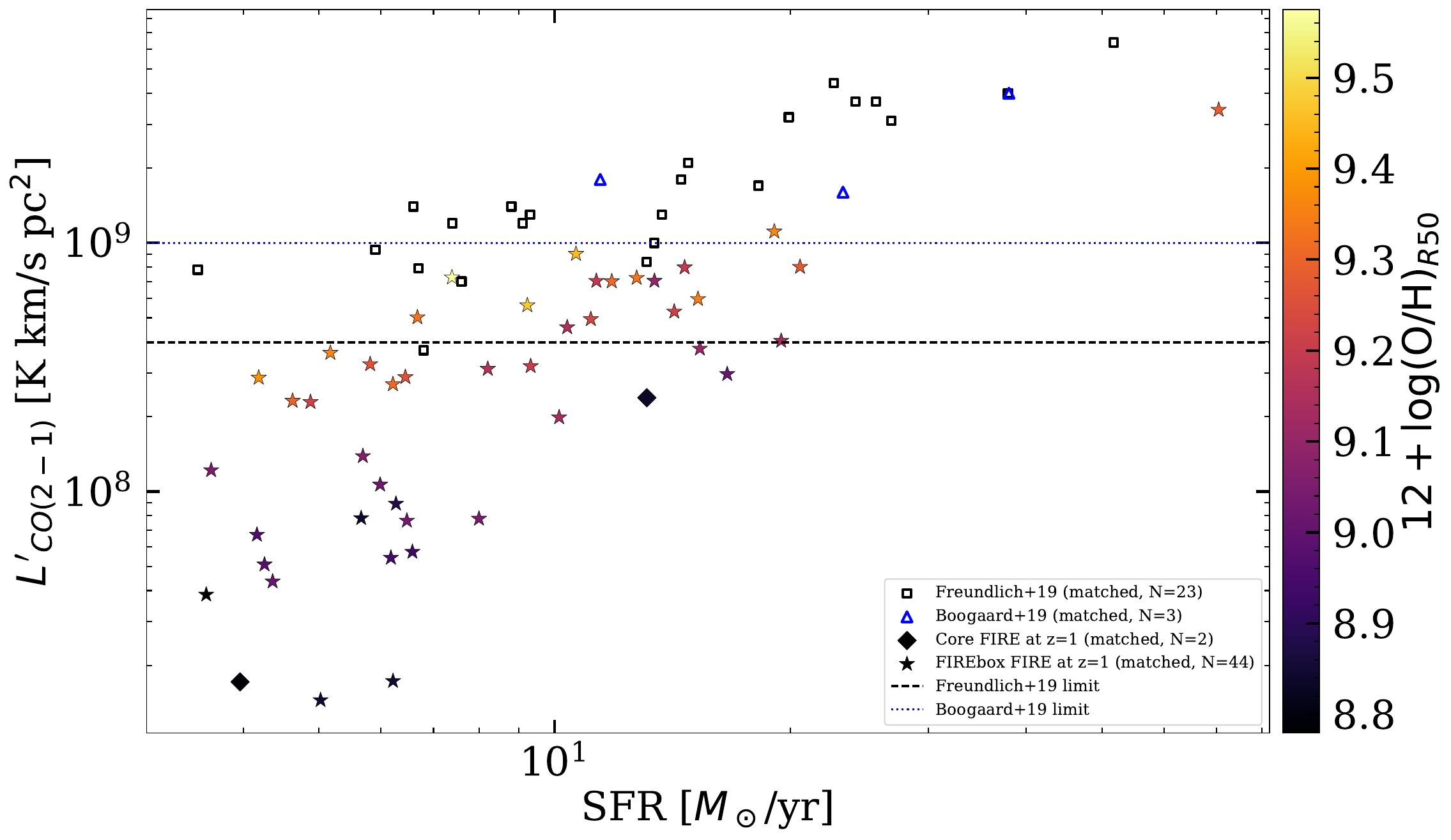}
    \caption{CO(2--1) luminosity versus SFR for the matched galaxies of Fig.~\ref{fig:matched_fgas_tdep_vs_Mstar_z1_sfr100Myr_Mstar0.2_fgas0.2_tdep0.2}, i.e.\ \fire\ and observed galaxies with similar $\Mstar$, molecular gas fraction and depletion time (all within 0.2\,dex). Open black squares and blue triangles show the PHIBSS2 \citep{Freundlich_2019} and ASPECS \citep{boogaard_2019_alma_aspecs} galaxies. Stars (\firebox) and diamonds (\core\ \fire) show the \fire\ galaxies at $z = 1$, coloured by their gas-phase oxygen abundance, $\twelvePlusOH_{R50}$, within the half-light radius $R_{50}$. PHIBB2 observations does not have measured metallicities and only handful of ASPECS galaxies have measured metallicities, so we do not show their metallicities in the figure. Dashed and dotted lines mark the PHIBSS2 ($\log L'_{\rm lim} = 8.6$) and ASPECS ($\log L'_{\rm lim} = 9.0$) sensitivity limits. Even with matched molecular gas content, \fire\ galaxies are fainter in CO(2--1), by a median of $0.64$\,dex. Most of this offset comes from the higher CO-to-H$_2$ conversion factor in \fire\: $\alpha_{\rm CO} \propto (Z/Z_\odot)^{-1.45 \pm 0.16}$ is about 10 times the Galactic value at fixed metallicity. Most of the vertical scatter of the \fire\ galaxies comes from metallicity. For example at ${\rm SFR} < 10\,M_\odot\,{\rm yr}^{-1}$, $L'_{\rm CO(2-1)} \propto (Z/Z_\odot)^{2.0}$ with no significant residual dependence on SFR. Only the most metal-rich \fire\ galaxies, $\twelvePlusOH_{R50} \gtrsim 9.3$, reach the CO luminosities of the observed galaxies}
    \label{fig:matched_Lco21_vs_sfr100Myr_z1_Mstar0.2_fgas0.2_tdep0.2}
\end{figure*}

% In this section we do a more controlled comparison between the observational dataset at moderate-redshift from PHIBBS \citep{Freundlich_2019} and ASPECS \citep{boogaard_2019_alma_aspecs} with our \fireTwo\ predictions by only comparing the galaxies at the same stellar mass, SFR, gas fraction and depletion time range. 

In this section we perform a more controlled comparsion between our \fireTwo\ predictions and the moderate redshift observations from PHIBBS2 \citep{Freundlich_2019} and ASPECS \citep{boogaard_2019_alma_aspecs}. To do so, we restrict both samples to galaxies that occupy the same range in stellar mass, SFR, gas fraction and depletion times. 

Figure~\ref{fig:matched_fgas_tdep_vs_Mstar_z1_sfr100Myr_Mstar0.2_fgas0.2_tdep0.2} shows the observed and simulated galaxies included in this comparsion. The gas fraction and depletion time are defined as: 
\begin{equation}
    \rm f_{gas} = \frac{M_{mol}}{M_{mol} + \Mstar}
\end{equation}
and 
\begin{equation}
    \rm t_{dep}/Gyr = \frac{M_{mol}}{SFR} 
\end{equation}
where $\rm M_{mol}$ is the molecular gas mass, including the contribution from helium. The observational and simulated estimates of $\rm M_{mol}$ are obtained very differently. The observers first convert their CO(2-1) luminosities to CO(1-0) by assuming a CO-to-$\rm H_2$ conversion factor, $\alpha_{CO}$. Each of these assumptions carries significant systematic uncertainties on its own. For our simulated galaxies, by contrast, we compute the molecular hydrogen mass $\rm M_{H_2}$ computed from running \cloudy\ \citep{cloudy_version_c25_release}, following the method described in \citetalias{Paper1}.

Only a small fraction of the observed galaxies have gas fractions and depletion times similar to those of our \fireTwo\ galaxies. We consider an observed galaxy to be matched it it lies within 0.2 dex of at least one \fireTwo\ galaxy in both $\rm f_{gas}$ and $\rm t_{dep}$. Under this criterion, 23 of the 55 PHIBSS2 and 3 of the 11 ASPECS galaxies are retained. Likewise, we kept 46 of the 76 \fireTwo\ galaxies, since the remaining 30 lie more than 0.2 dex from any observed galaxy. The final matched sample therefore consists of 46 \fireTwo\ galaxies paired with 26 observed galaxies. Each observed galaxy has between 1 to 12 \fireTwo\ counterparts, with a median of 3.5.

Figure~\ref{fig:matched_Lco21_vs_sfr100Myr_z1_Mstar0.2_fgas0.2_tdep0.2} shows \LcoTwoOne\ as a function of SFR for this matched sample. The colorbar shows the oxygen abundance of the \fireTwo\ galaxies within \RvisibleHl, $\twelvePlusOH_{R50}$. We do not show the metallicities of the observations, as they are either inferred from the mass-metallicity relation as in the case of PHIBBS2 or only one of the three galaxies in ASPECS have metallicities recorded. 

The \fireTwo\ predictions have a scatter of approximately 0.53 dex nearly twice the 0.30 dex scatter of the observations. The \fireTwo\ galaxies also lie systematically below the observations, with a median offset by a factor of 5 that shows no dependence on SFR. This offset is driven mainly by higher $\alpha_{CO}$ values of the \fireTwo\ galaxies relative to those assumed in the observational studies. The \fireTwo\ galaxies with $\LcoTwoOne < 10^8\,\obsunits$ have a median $\alpha_{CO} \approx 32\,\rm M_\odot\,(K\,km\,s^{-1}\,pc^2)^{-1}$, nearly an order of magnitude above the value adopted by the observations. In contrast, those with $\LcoTwoOne > 4\times10^{8}\,\obsunits$ have $\alpha_{CO} \approx 5.3$, close to the observational value. 

This systematic difference in $\alpha_{CO}$ arises from the metallicities of the \fireTwo\ galaxies. For the plotted \fireTwo\ galaxies $\log(\alpha_{CO})$ decreases with $\twelvePlusOH_{R50}$ with a slope of $\rm d\log(\alpha_{CO})/d(\twelvePlusOH_{R50})=-1.45$, a superlinear dependence on metallicity. Therefore, with decreasing metallicity the amount CO dark gas increases at the same SFR. 

Metallicity can also account for much of the vertical spread in the \fireTwo\ predictions. Among the \fireTwo\ galaxies in this figure with $\rm SFR < 10\,M_\odot\,yr^{-1}$, $\log(\LcoTwoOne)$ increases with $\twelvePlusOH_{R50}$ with a slope of about 2. Because these galaxies span a metallicity range of about 0.8 dex, this dependence alone accounts for a vertical spread of roughly 1.6 dex in \LcoTwoOne.

\clearpage
\bibliography{bibliography}{}
\bibliographystyle{aasjournal}

\end{document}